\documentclass[prb,amsmath,amssymb,lengthcheck,showpacs,floatfix,groupedaddress,superscriptaddress,longbibliography]{revtex4-1}
\usepackage{graphicx}
\usepackage[english]{babel}
\usepackage{times}
\usepackage{dcolumn}
\usepackage[usenames,dvipsnames]{color}
\usepackage{units}
\usepackage{bm}

\begin{document}

\title{Spectral properties of Shiba sub-gap states at finite temperatures}

\author{Rok \v{Z}itko}

\affiliation{Jo\v{z}ef Stefan Institute, Jamova 39, SI-1000 Ljubljana, Slovenia}
\affiliation{Faculty  of Mathematics and Physics, University of Ljubljana, 
Jadranska 19, SI-1000 Ljubljana, Slovenia}

\date{\today}

\begin{abstract}
Using the numerical renormalization group (NRG), we analyze the
temperature dependence of the spectral function of a magnetic impurity
described by the single-impurity Anderson model coupled to
superconducting contacts. With increasing temperature the spectral
weight is gradually transferred from the $\delta$-peak
(Shiba/Yu-Shiba-Rusinov/Andreev bound state) to the continuous sub-gap
background, but both spectral features coexist at any finite
temperature, i.e., the $\delta$-peak itself persists to temperatures
of order $\Delta$. The continuous background is due to inelastic
exchange scattering of Bogoliubov quasiparticles off the impurity and
it is thermally activated since it requires a finite thermal
population of quasiparticles above the gap. In the singlet regime for
strong hybridization (charge-fluctuation regime) we detect the
presence of an additional sub-gap structure just below the gap edges
with thermally activated behavior, but with an activation energy equal
to the Shiba state excitation energy. These peaks can be tentatively
interpreted as Shiba bound states arising from the scattering of
quasiparticles off the thermally excited sub-gap doublet Shiba states,
i.e., as high-order Shiba states.
\end{abstract}


\maketitle

\newcommand{\vc}[1]{{\mathbf{#1}}}
\renewcommand{\Im}{\mathrm{Im}}
\renewcommand{\Re}{\mathrm{Re}}

\newcommand{\expv}[1]{\langle #1 \rangle}
\newcommand{\ket}[1]{| #1 \rangle}
\newcommand{\Tr}{\mathrm{Tr}}

\section{Introduction}

A magnetic impurity in a superconducting host induces localized bound
states inside the spectral gap, known in different communities as
either Shiba, Yu-Shiba-Rusinov, or Andreev bound states
\cite{shiba1968,shiba1973,satori1992,balatsky2006,yazdani1997,Deacon:2010jn,pillet2010,franke2011,rodero2011}.
At zero temperature, Shiba states manifest as pairs of $\delta$-peak
resonances in the impurity spectral function $A(\omega)$ positioned
symmetrically at positive and negative frequency corresponding to the
transitions from the many-particle ground state to the {\sl same}
many-particle excited state by either adding a probing electron to the
system $(\omega>0)$ or removing it $(\omega<0)$. The intrinsic
temperature dependence of the spectral function depends on the
impurity dynamics. When the impurity behaves as a classical object,
i.e., a local magnetic field which is perfectly {\sl static} on the
time-scale of the experiment (``adiabatic limit'' with no dynamics of
the internal degrees of freedom of the impurity), the corresponding
{\sl classical} impurity model is a quadratic non-interacting
Hamiltonian, hence the spectral function is not temperature dependent
at all. This problem can be discussed in terms of single-particle
levels and their occupancy. When the impurity behaves, however, as a
quantum object, i.e., a {\sl fluctuating} local moment as described by
the Kondo or Anderson {\sl quantum} impurity models, there will be
non-trivial intrinsic temperature dependence due to electron-electron
interactions (inelastic exchange scattering of thermally excited
Bogoliubov quasiparticles off the impurity spin). This problem is
better addressed from the perspective of many-particle eigenstates.
Since the eigenvalue spectrum of the Hamiltonian operator includes
both discrete Shiba states below the gap and a continuum part above
the gap, it is expected that there will be both $\delta$-peaks and a
continuous background coexisting inside the gap at any finite
temperature, providing a further realisation of the ``bound state in
the continuum'' paradigm.

The temperature dependence of the Andreev spectra was studied
experimentally in carbon nanotube quantum dots \cite{Kumar:2014cq}.
Strong temperature effects found in the measured differential
conductance could be accounted for reasonably well using the tunneling
formalism, however the intrinsic temperature dependence of the
impurity spectral function was not discussed. Another experimental
realization of impurity models are magnetic adatoms on superconducting
surfaces. In Ref.~\onlinecite{ruby2015} the measured differential
conductance was discussed in terms of a phenomenological impurity
model based on a classical impurity. There is, however, a lack of
theoretical works on the temperature dependence of spectra of quantum
impurity models to provide an alternative framework from the
interpretation of measured spectra.

In this work we study the sub-gap spectral features in the
single-impurity Anderson model with a superconducting bath described
by the $s$-wave BCS mean-field Hamiltonian. After introducing the
model and methods in Sec.~\ref{sec2}, we first consider the model by
fixing the gap parameter $\Delta$ to its zero-temperature value and
increasing the temperature $T$ in Sec.~\ref{sec3}. This simplified
calculation uncovers how the spectral weight is transferred from the
Shiba $\delta$-peak to the continuum. In this section we also study
the hybridization dependence and the differences between the singlet
(screened impurity) and doublet (unscreened impurity) regimes. In
Sec.~\ref{sec4} we perform a full calculation with the temperature
dependent gap of a BCS superconductor; in this case the Shiba peak
broadening is accompanied by peak shifts. We conclude with a
discussion of the experimental relevance of the results.

\section{Model and method}
\label{sec2}

We consider the Hamiltonian $H = H_\mathrm{BCS} + H_\mathrm{imp} +
H_\mathrm{c}$:
\begin{equation}
\begin{split}
H_\mathrm{BCS} &= \sum_{k\sigma} \epsilon_k c^\dag_{k\sigma} c_{k\sigma}
- \Delta \sum_k \left( c^\dag_{k\uparrow} c^\dag_{k\downarrow} + \text{H.c.}
\right), \\
H_\mathrm{imp} &= \epsilon_d \sum_\sigma n_\sigma + U n_\uparrow
n_\downarrow, \\
H_\mathrm{c} &= \sum_{k\sigma} V_k \left( c^\dag_{k\sigma} d_\sigma +
\text{H.c.} \right).
\end{split}
\end{equation}
Here $c_{k\sigma}$ and $d_\sigma$ are the band and impurity electron
annihilation operators, $\epsilon_k$ the band dispersion relation,
$\Delta$ the BCS gap parameter, $\epsilon_d$ the impurity level, $U$
the electron-electron repulsion, $n_\sigma=d^\dag_\sigma d_\sigma$ the
impurity occupancy operator, and $V_k$ the hopping integrals.  The
Hamiltonian does not include any coupling to electromagnetic noise or
phonons.

Assuming a flat band with the density of states $\rho$ in the normal
state, and $V_k \equiv V$, the impurity coupling is fully
characterized by the hybridization strength $\Gamma=\pi \rho V^2$. In
this work, we focus on the particle-hole (p-h) symmetric case with
$\epsilon_d=-U/2$. The Kondo exchange coupling at $\Delta=0$ is given
by the Schrieffer-Wolff transformation as $\rho J_K = 8\Gamma/\pi U$
and the Kondo temperature as \cite{krishna1980a}
\begin{equation}
T_K^0 \sim U \sqrt{\rho J_K} \exp \left( -\frac{1}{\rho J_K} \right).
\end{equation}

\begin{figure}
\centering
\includegraphics[clip,width=0.48\textwidth]{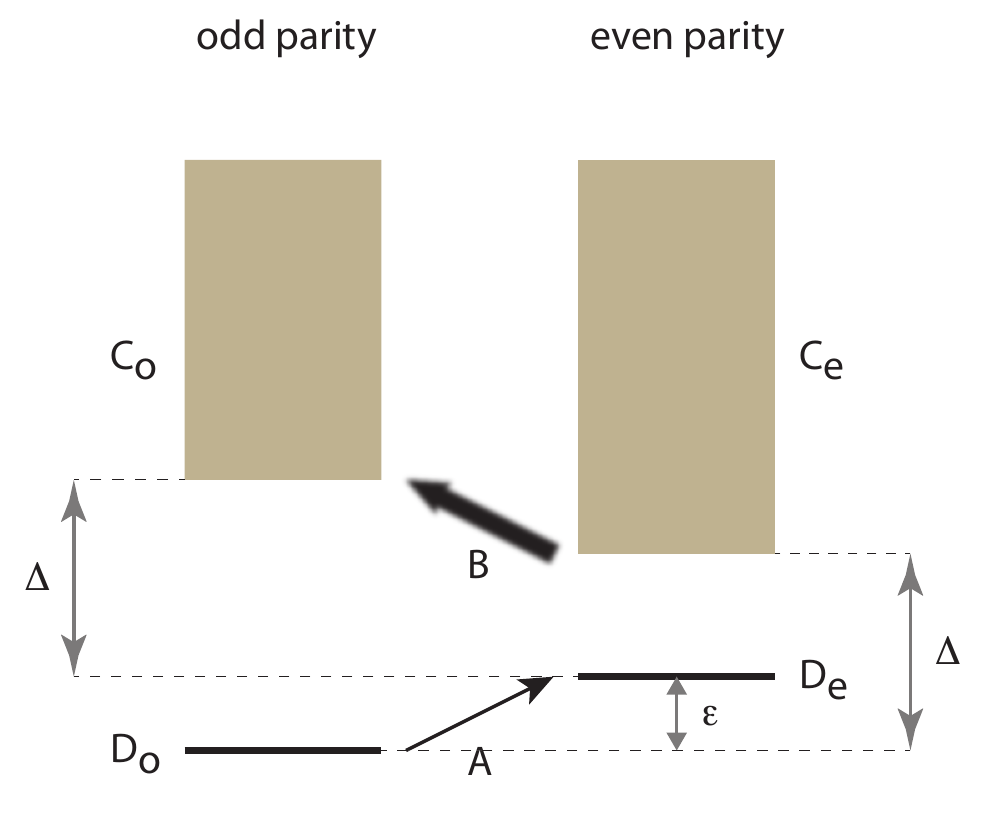}
\caption{(Color online) Schematic diagram of the many-particle
eigenstates of the Hamiltonian, partitioned into the even and odd
fermion-parity sectors (i.e., parity of the total electron number).
This diagram corresponds to the case where the ground state has odd
parity (spin-doublet). $D_o$ and $D_e$ are the odd-parity
(spin-doublet) $\ket{D}$ and the even-parity (spin-singlet) $\ket{S}$
discrete eigenstates. The even-parity continuum $C_e$ starts at energy
$\Delta$ above the odd-parity discrete state $D_o$, since the
bottom-most states of the continuum are composed of one additional
quasiparticle added to $D_o$, thus changing the overall fermion
parity. The odd-parity continuum $C_o$ starts at energy $\Delta$ above
the even-parity discrete state $D_e$, for similar reasons. Note that
the multiple-quasiparticle states have energies at least $2\Delta$
above $D_o$. Label $A$ indicates a sharp transition (contributing a
$\delta$-peak to the impurity spectrum), label $B$ diffuse
transitions (contributing a continuous background to the spectrum). }
\label{fig1}
\end{figure}

In the superconducting case with $\Delta \neq 0$, the ground state of
the system is either a singlet $\ket{S}$ or a doublet $\ket{D}$
depending on the value of the ratio $\Delta/T_K^0$. All other
eigenstates are, in the first approximation (i.e., neglecting residual
interactions between the quasiparticles), product states of either
$\ket{S}$ or $\ket{D}$ with additional Bogoliubov quasiparticles from
the continuum. While the total particle number is not a conserved
quantum number for $\Delta \neq 0$, its parity is. The eigenstates can
thus be classified into odd and even fermion parity sectors, as
illustrated for the case of an odd-parity (spin doublet) ground state
in Fig.~\ref{fig1}. A quasiparticle is an object with odd
fermion-parity, thus the even-parity continuum starts at the energy
$\Delta$ above the odd-parity ground state, while the odd-parity
continuum starts at the energy $\epsilon+\Delta$ above the ground
state, where $\epsilon$ is the Shiba state ``energy'' (more precisely,
the energy difference
\begin{equation}
\epsilon=|E_S-E_D|
\end{equation}
between the sub-gap many-particle Shiba states).

In this work we are interested mainly in the spectral functions at
finite $T$. The calculations are performed with the numerical
renormalization group (NRG)
\cite{wilson1975,satori1992,sakai1993,yoshioka2000,oguri2004josephson,bauer2007,karrasch2008,bulla2008,zitko2015shiba}.
This method appears at first perfectly suited for the problem, since
it is an unbiased nonperturbative numerical technique, applicable both
at zero and at finite temperatures, which can handle arbitrary bath
density of states (including with a superconducting gap), and provides
the spectral function directly on the real frequency axis. Other
methods are either biased, perturbative, inapplicable to the
superconducting case, or require an analytical continuation from the
Matsubara axis to real frequencies; in particular, this last issue
makes the quantum Monte Carlo (QMC) approach of little use, since it
is extremely difficult to perform an analytical continuation in the
presence of a sharp gap, especially since it is necessary (see below)
to resolve a $\delta$-peak superposed on a continuous background of
finite support inside the gap. Nevertheless, the situation under study
in this work is in some regards perhaps the worst possible case for
the NRG. While the method works very well for problems with spectral
gap at zero temperature, and for non-gapped baths at any temperature,
there are severe difficulties when both $\Delta$ and $T$ are non-zero.
Both the gap and the temperature break the scale invariance on which
the method is based, and they do so in different ways, thereby
generating inevitable systematical errors. The results for spectral
functions presented in this work should thus be considered as
qualitatively correct, while quantitative errors are estimated (by
monitoring how the results fluctuate when the NRG calculation
parameters are varied) to be in the tens of percent range for
$T\sim\Delta$. In spite of this shortcoming, there is presently no
other impurity solver to meaningfully study the finite-temperature
spectral function. Static properties, such as the expectation values
of various operators, can be reliably computed using the QMC
\cite{Luitz:2010bn,gull2011}. Even here, there are some small
systematic discrepancies between the QMC and NRG when {\sl both}
$\Delta$ and $T$ are non-zero. Such comparisons of static properties
are very useful to tune the parameters of the NRG to values where such
discrepancies are minimal. Finally, we note that the
finite-temperature problems in the NRG become severe when $U$ is
small, while they seem to be more manageable in the deep Kondo regime
which is of main interest in this study.

The NRG calculations were performed with the discretization parameter
$\Lambda=2$, with $N_z=8$ interleaved discretization grids
\cite{oliveira1994,resolution}, using the full-density-matrix
algorithm with the Wilson chain terminated at the energy scale
$E_\mathrm{chain}=\Delta/50$
\cite{anders2005,peters2006,weichselbaum2007}. The ``traditional''
choice of the discretization parameter $\Lambda=2$ proved to be near
optimal. The results depend little on the choice of the discretization
method \cite{resolution}. The length of the Wilson chain, however,
turned out to be a critical parameter and had to be tuned.

To obtain a good description of the continuum part of the sub-gap
spectrum at finite temperatures, it furthermore proved crucial to keep
a large number of states in the NRG iteration even at energy scales
below $\Delta$, much more than required for obtaining well converged
thermodynamics and $T=0$ spectral functions; we kept at least 2500
multiplets. While computationally demanding, this is critically
important for a good description of the continuum quasiparticle
spectrum in both even- and odd-parity parts of the full Fock space.
\footnote{A possible improvement consists in formulating the NRG
truncation rule so that a comparable total number of multiplets is
kept in even- and odd-fermion-parity sectors, but this has not yet
been tried out.}

The impurity Green's function is defined as
\begin{equation}
G(t) = -i\theta(t) \Tr\left\{ \rho [d_\sigma(t), d_\sigma^\dag(0) ]_+ \right\},
\end{equation}
where the trace is evaluated with the grand-cannonical density matrix
$\rho=e^{-\beta H}$ (the chemical potential is fixed to $\mu=0$). This
is an appropriate description only for well equilibrated ergodic
systems. The assumption of ergodicity is non-trivial and may not be
valid in all impurity systems and under all experimental conditions.
Furthermore, the presence of the tunneling contacts will drive the
system out of equilibrium. 

The impurity spectral function, 
\begin{equation}
A(\omega)=-\frac{1}{\pi} \Im {\tilde G}(\omega+i\delta),
\end{equation}
where ${\tilde G}$ is the Fourier transform of $G$, can be expressed
using the Lehmann decomposition as
\begin{equation}
\begin{split}
A(\omega) = & \frac{1}{Z} \sum_{mn} \left| \left\langle
m | d_\sigma | n \right\rangle \right|^2 \times\\
&
\times \left( e^{-\beta E_m} + e^{-\beta E_n} \right) \delta(\omega+E_m-E_n),
\end{split}
\end{equation}
where $m,n$ index all eigenstates of the Hamiltonian, $E_{m,n}$ are
the corresponding eigenvalues, $\beta=1/k_B T$, and the
grand-cannonical partition function is $Z=\Tr[\exp(-\beta H)]=\sum_m
\exp(-\beta E_m)$. The actual calculation of $A(\omega)$ is performed
using the full-density-matrix algorithm \cite{weichselbaum2007},
generalizing the complete-Fock-space approach
\cite{anders2005,peters2006}. We accumulate the raw spectral data
separately for $|\omega| < \Delta$ and $|\omega|>\Delta$. Inside
the gap, we use 5000 equidistant bins. Outside the gap, we use a
logarithmic mesh of bins with low-frequency accumulation points at
$\omega = \pm \Delta$ and with 1000 bins per frequency decade. This
modification of the standard binning is necessary for obtaining
constant spectral resolution inside the gap and a correct description
of the gap edges in the continuum above the gap \cite{hecht2008}.

The Green's function probes the single-particle excitations of the system.
It should be emphasized that all contributions to $G$ correspond to
electron-parity-changing transitions (see Fig.~\ref{fig1}). Let us consider 
the doublet regime, where the impurity spin is
unscreened and the ground state is the odd-parity spin-doublet
$\ket{D}$. At zero temperature, only the ground state $D_o$ is
thermally occupied, and the only transition with $\Delta E < \Delta$
is that to the discrete excited state $D_e$ (transition $A$ indicated
by the sharp arrow in Fig.~\ref{fig1}). The sub-gap part of the
spectrum is thus fully described by two $\delta$-peaks at positions
$\omega=\pm \epsilon$ with equal weight (due to the p-h symmetry)
given by
\begin{equation}
w_\delta(T=0) = \frac{1}{2} \left| \left\langle D_o | d_\sigma | D_e \right\rangle \right|^2.
\end{equation}
At finite temperatures there are further transitions with starting
and end states separated by less than $\Delta$: they are indicated by
a diffuse arrow $B$ in Fig.~\ref{fig1} and correspond to transitions
from the thermally populated even-parity quasiparticle states at
energies above $\Delta$ (set $C_e$) to the odd-parity quasiparticle
states at energies above $\epsilon+\Delta$ (set $C_o$). Since the
states involved form continua, this will generate a continuous
spectral weight contribution to the sub-gap spectrum. The most likely
transitions are those from the bottom of $C_e$ to the bottom of $C_o$,
thus the continuum background is expected to be peaked at
$|\omega|=\epsilon$, i.e., at the position of the discrete Shiba state
which itself persists at finite temperature at least up to $T \sim
\Delta$. The evolution with increasing $T$ is thus expected to be as
follows: the weight of the $\delta$-peak decreases, while the weight
of a new broad peak centered at the same position increases. In the
limit of high temperatures, $T \gg \Delta$, the partition function $Z$
is large, the discrete contribution $A$ to the spectral function is
negligible and there is only continuum weight (this happens in
the $T \to T_c$ limit, where $\Delta(T) \to 0$). In the next sections,
we confirm this intuitive physical picture by numerical calculations.

\section{Results: fixed $\Delta$}
\label{sec3}

\subsection{Overview and main characteristics}

The calculations in this subsection are performed for fixed
model parameters ($\Gamma/U=0.1$ and $U/\Delta=20$), only the temperature $T$ is
variable. The ground state is a spin doublet, while the singlet
excited state lies at the energy level
\begin{equation}
\epsilon=0.423 \Delta
\end{equation}
above it. Due to the p-h symmetry the spectral function is even and we
focus on its $\omega>0$ (particle addition) part. At zero temperature,
the weight of the $\delta$-peak at $\omega=\epsilon$ is
\begin{equation}
w_\delta(T=0)=0.0341.
\end{equation}
This indicates that the Shiba bound state wavefunction (as far as it
can be defined for an interacting system) has majority of its weight
not on the impurity, but in the host, which is commonly the case for
Shiba states.

At finite temperatures some care is required in post-processing the
raw spectral data as obtained from the NRG run. The
$\delta$-peak is extracted from the spectral function by removing the weight in a
narrow interval of width $2\times 10^{-4} \Delta$ around
$\omega=\epsilon$, where $\epsilon$ can be independently determined very
accurately from the NRG flow diagrams. The remaining continuous part of the 
spectral function is then broadened and further
characterized. This procedure allows us
to reliably partition the spectral function into discrete and
continuous components:
\begin{equation}
A(\omega)=A_\delta(\omega)+A_c(\omega).
\end{equation}
The corresponding spectral weights are defined as 
\begin{equation}
w_i = \int_0^\Delta A_i(\omega) d\omega
\end{equation}
with $i=\delta,c$.  It should be noted in passing that at finite
temperatures $w_\delta(T)$ receives contributions not only from the
transition A, but also from a discrete subset of transitions between
the states forming the continua $C_e$ and $C_o$ with energy difference
{\sl exactly} equal to $\epsilon$ (i.e., the transitions $D_e \to D_o$
in the presence of quasiparticles, but without the quasiparticles
interacting with the impurity). The temperature dependence of both
$w_\delta$ and $w_c$ is an interaction effect: for a non-interacting
Hamiltonian, such as that corresponding to a classical impurity with
no internal dynamics, the spectral function itself would not depend in
any way on the temperature (although the {\sl occupancies} of the
single-particle levels would change with $T$). 

\begin{figure}
\centering
\includegraphics[clip,width=0.48\textwidth]{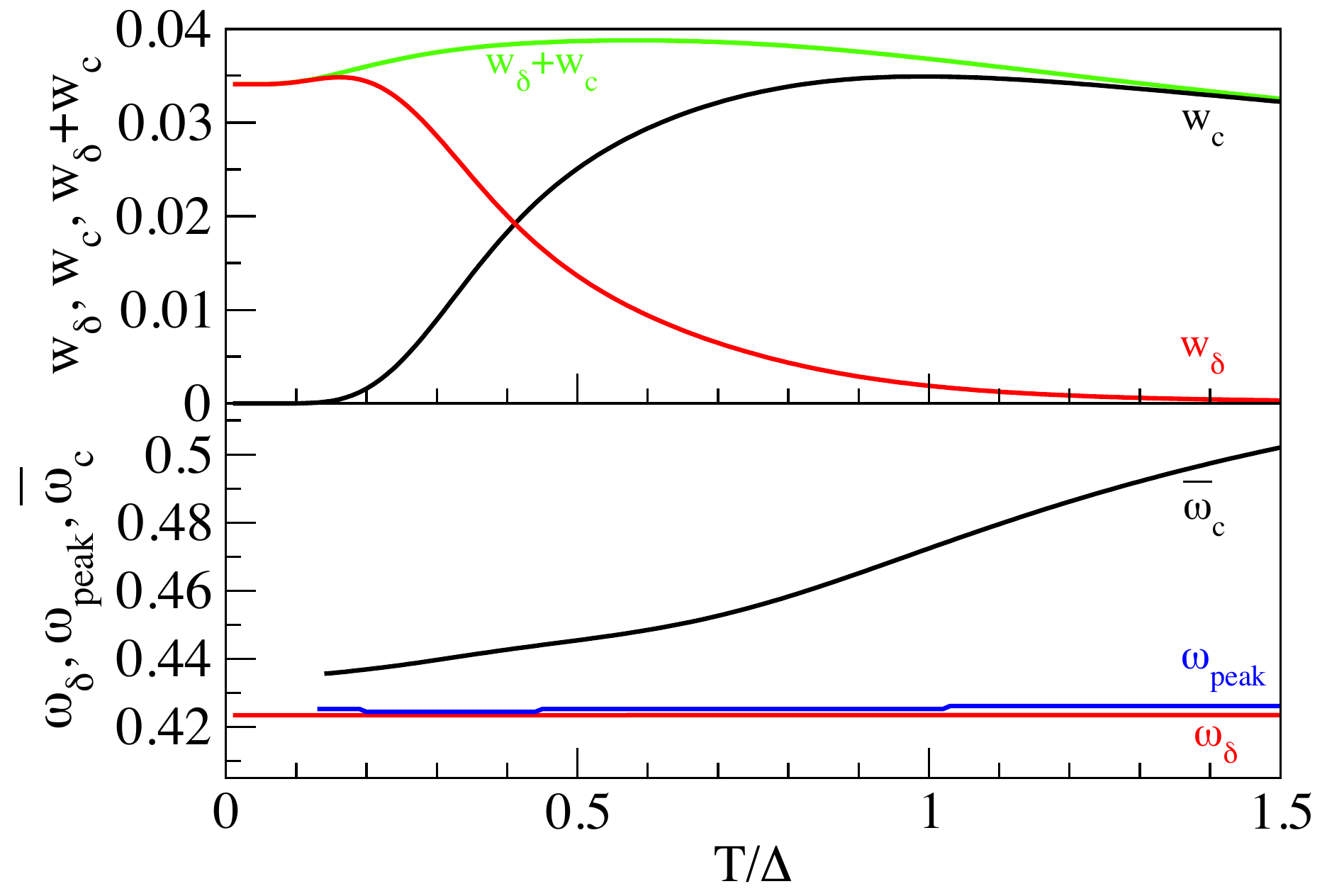}
\caption{(Color online) (a) $\delta$-peak, continuum, and total
spectral weight in the positive-frequency sub-gap part
($0<\omega<\Delta$) of the impurity spectral function $A(\omega,T)$.
(b) Position of the $\delta$-peak, $\omega_\delta$, and of the maximum
of the continuous part, $\omega_\mathrm{peak}$, as well as the mean
value of the continuous part, ${\bar \omega}_c$.
}
\label{fig2}
\end{figure}

\begin{figure}
\centering
\includegraphics[clip,width=0.48\textwidth]{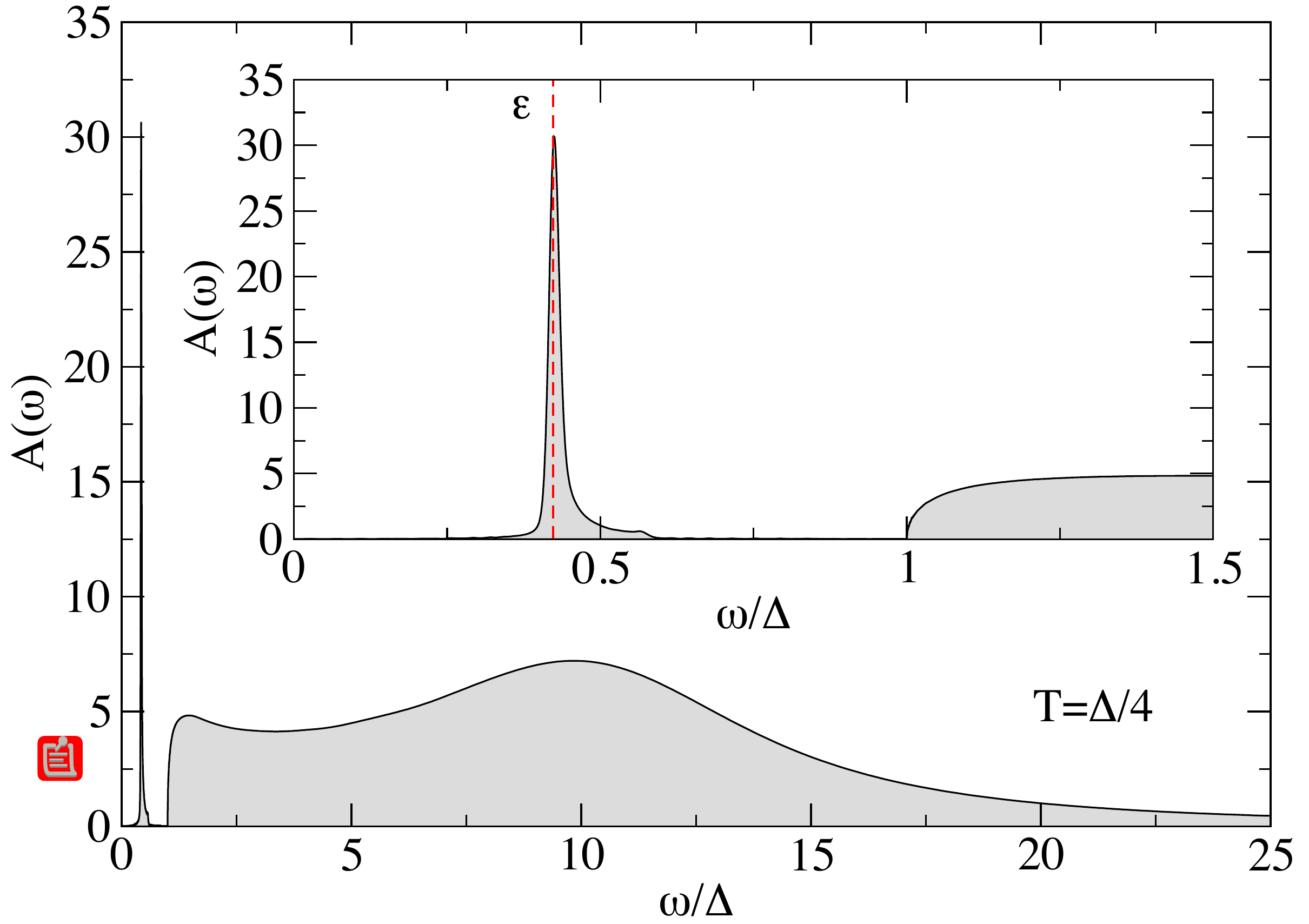}
\caption{(Color online) Impurity spectral function $A(\omega,T)$ at
finite temperature $T=\Delta/4$. The hump at $\omega=10\Delta=U/2$ is
the Hubbard peak. The inset shows a close-up on the sub-gap region.
The position of the $\delta$-peak, $\epsilon$, is indicated using the
dashed line and essentially coincides with the peak of the continuum
part.}
\label{spec3}
\end{figure}

The most important spectral characteristics are revealed in the
temperature-dependence plots shown in Fig.~\ref{fig2}, while an
example of a typical finite-$T$ spectral function is shown in
Fig.~\ref{spec3}.

The continuum weight $w_c$ exhibits activated behavior for low $T$, with the
activation energy $\Delta$:
\begin{equation}
w_c(T) = 0.168\, e^{-\Delta/T}.
\end{equation}
This confirms the expectation that the continuum background is
associated with the {\sl inelastic} transitions that require a finite
thermal population of the quasiparticle states above the gap which
scatter on the impurity (diffuse transitions as shown schematically in
Fig.~\ref{fig1}, arrow B). 

For $T \gtrsim 0.2 \Delta$, $w_\delta$ is a strictly decreasing
function of temperature, while $w_c$ is increasing, and their sum
$w_\delta+w_c$ is approximately constant: the weight is gradually
transferred from the coherent discrete Shiba state to diffuse states
involving itinerant quasiparticle states, i.e., this represents a
thermal decomposition of the Shiba state. We note that $w_\delta =
w_c$ on the scale $T \approx \Delta/2$. This is also the range where
the total weight $w_\delta+w_c$ reaches a maximum value. The continuum
weight $w_c$ is increasing up to $T \approx \Delta$ where it reaches a
value close to $w_\delta(T=0)$. In simple terms, with increasing
temperature almost all spectral weight is transferred from the
$\delta$-peak to the continuum by $T \approx \Delta$. For $T>\Delta$,
$w_c$ itself becomes a decreasing function, albeit only weakly: the
decay of $w_\delta$ at large $T$ is much faster than that of $w_c$,
and $w_\delta$ becomes essentially zero by $T \approx 2\Delta$.

In Fig.~\ref{fig2}(b) we consider the peak positions. The
$\delta$-peak does not move with temperature. This is expected, since
its position $\omega_\delta=\epsilon$ is given by the energy
difference of the two discrete eigenstates of the Hamiltonian, thus it
is a property of the operator itself and has nothing to do with
thermal effects. The continuum part of the sub-gap spectrum is a
peaked function, see Fig.~\ref{spec3}. The position of this peak,
$\omega_\mathrm{peak}$, almost coincides with the $\delta$-peak
position:
\begin{equation}
\omega_\mathrm{peak} \approx \omega_\delta = \epsilon.
\end{equation}
$\omega_\mathrm{peak}$ is very weakly temperature dependent, see
Fig.~\ref{fig2}(b). We also plot the mean of the continuum part,
${\bar \omega}_c$, defined as the normalized first moment of
$A_c(\omega)$. The mean is larger than $\epsilon$ and further
increases with $T$, indicating that the continuum part of the spectrum
is skewed toward larger frequencies, as can also be seen in
Fig.~\ref{spec3}. At low temperatures, the skewness exceeds 6. The
long tail is due to the asymmetry of the transitions: the most
populated thermally excited starting states are those near the bottom
of the even-parity continuum and most likely end states those at the
bottom of the odd-parity continuum starting at $\epsilon$ higher in
energies. At higher temperatures, $T \sim \Delta$, the distribution
becomes more symmetric around $\omega=\epsilon$ with a clear dominant
peak, corresponding to the ``thermally broadened'' Shiba resonance.

The width of the continuum part can be further characterized through
the standard deviation (not shown). It is a strictly increasing
function of $T$. At intermediate temperatures $T \approx \Delta/2$ it
reaches a value of order $0.1\Delta$, thus the background is
relatively broad. Another relevant quantity is the half-width at
half-maximum (HWHM) of the main peak in the continuum part. This
quantity is very difficult to extract reliably since it requires a
delicate broadening procedure and it strongly depends on the NRG
calculation parameters. We find that the HWHM is only weakly increasing in
the temperature range $T<\Delta$: it starts at values close to
$0.01\Delta$ in the low-temperature limit and increases to $\sim
0.015\Delta$ at $T=\Delta$. The main thermal effect is thus the weight
transfer from the discrete to the continuous part, but there appears to
be little broadening in the sense of decreasing lifetime of the
continuum resonance feature at $\omega=\epsilon$.

\subsection{$\Gamma$-dependence}

We now study how the results from the previous subsection depend on
the value of the hybridization $\Gamma$, in particular accross the
singlet-doublet quantum phase transition where $\ket{S}$ and $\ket{D}$
interchange their roles as the ground and the excited state,
respectively.

\begin{figure}
\centering
\includegraphics[clip,width=0.48\textwidth]{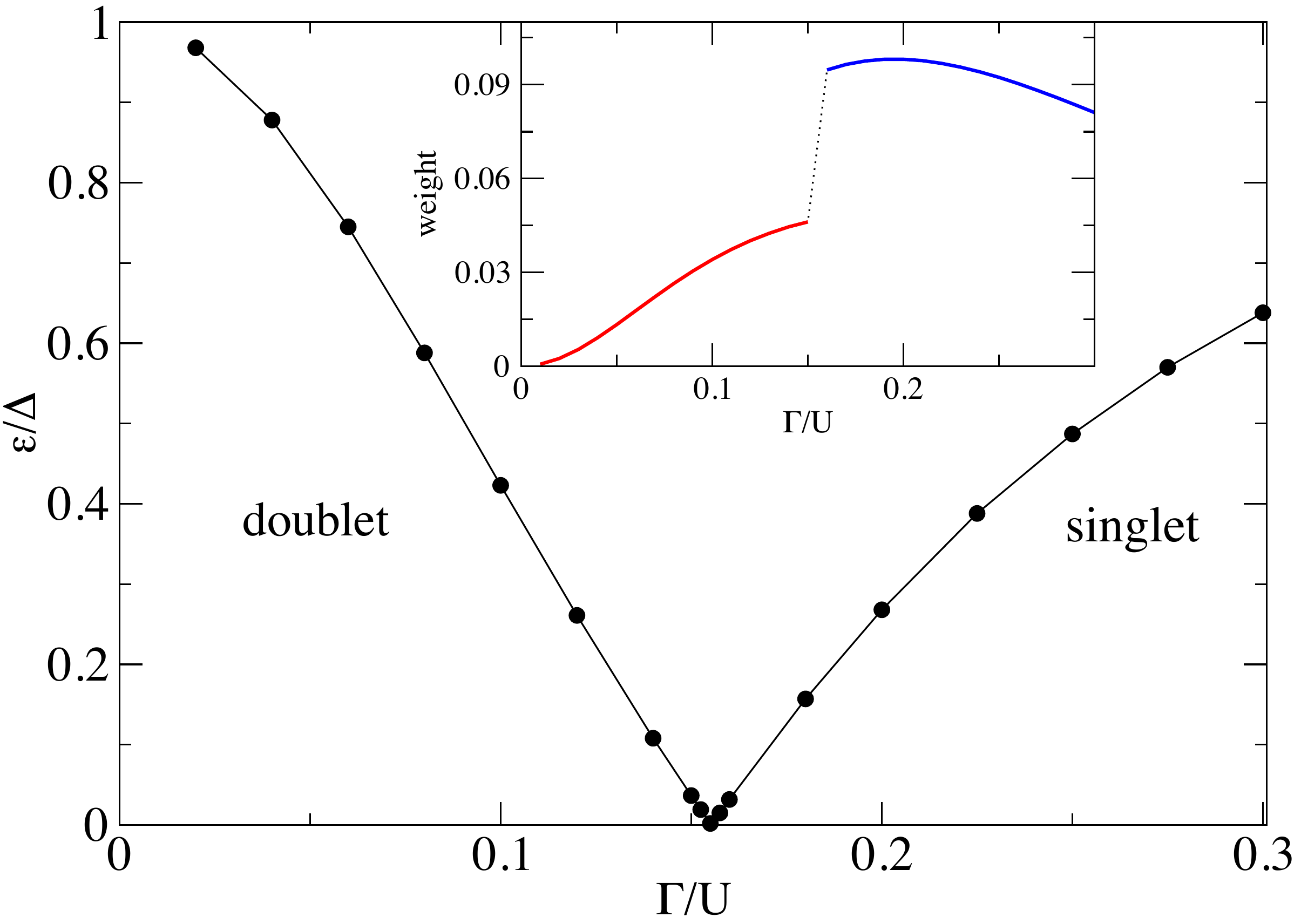}
\caption{(Color online) Shiba state energy $\epsilon$ as a function of the
hybridization strength $\Gamma$, for fixed $U/\Delta=20$. The inset
shows the $T=0$ spectral weight of the sub-gap $\delta$-peak.}
\label{figX}
\end{figure}

For low enough $\Gamma$, so that the impurity is in the Kondo regime,
the Shiba state energy $\epsilon$ follows the universal dependence
$\epsilon(T_K/\Delta)$, where $T_K=T_K(\Gamma)$. For $\Gamma \to 0$,
the peak is close to the gap edge, then it moves toward the chemical
potential for increasing $\Gamma$, see Fig.~\ref{figX}. For chosen
$U/\Delta=20$, the singlet-doublet (S-D) transition occurs at 
\begin{equation}
\Gamma_c =0.155U.
\end{equation}

\begin{figure}
\centering
\includegraphics[clip,width=0.48\textwidth]{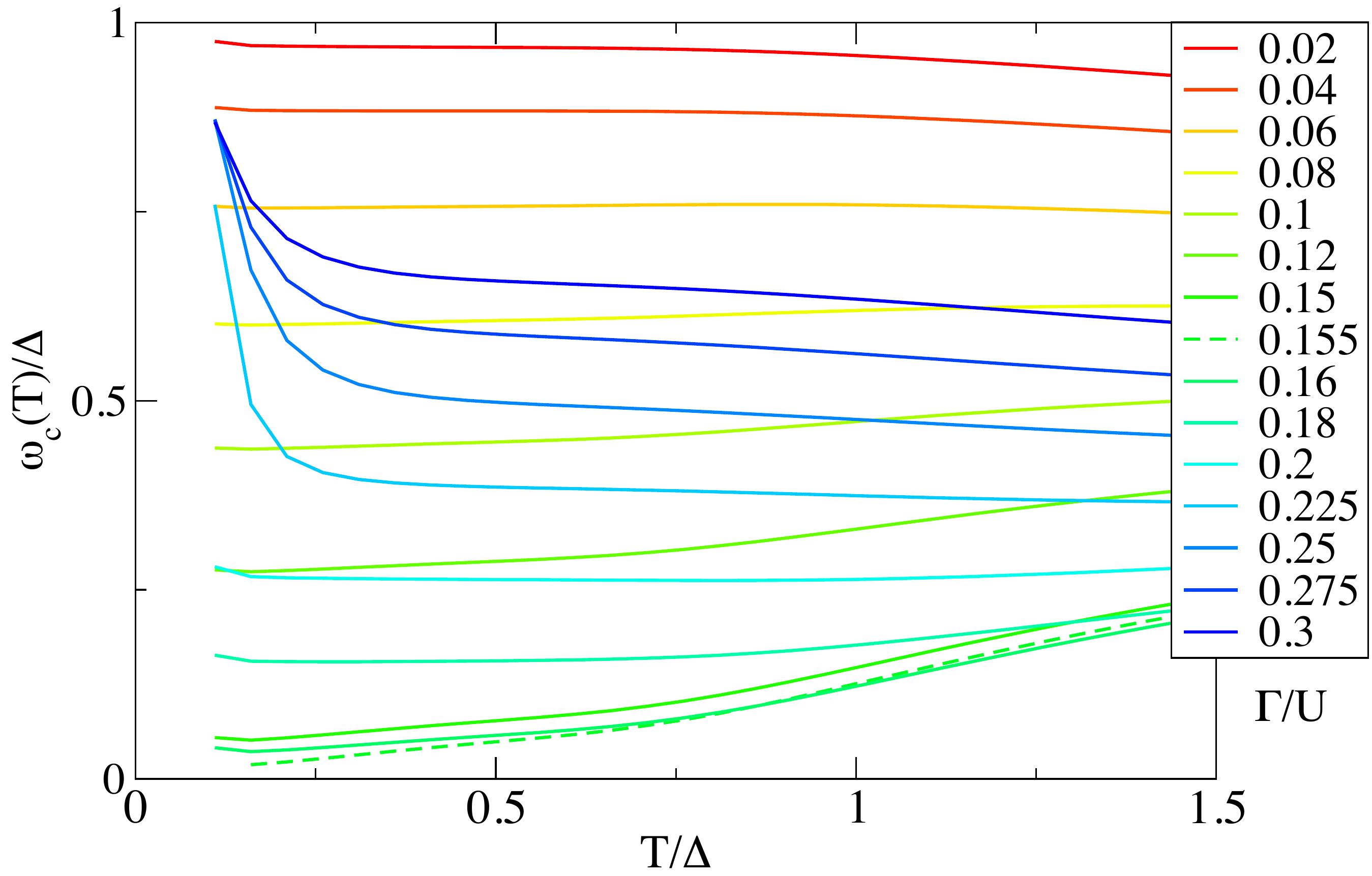}
\caption{(Color online) Temperature dependence of the mean-value of the
continuum part of the sub-gap spectrum, $\omega_c(T)$, for a range of
hybridization strengths $\Gamma$.}
\label{fig4a}
\end{figure}

We first consider how the temperature dependencies of the key spectral
characteristics change for different values of $\Gamma$. The
$\delta$-peak position $\omega_\delta=\epsilon$ does not vary with
temperature. The continuum mean, $\omega_c$, shown in
Fig.~\ref{fig4a}, starts from $\omega_c(T=0) \approx \epsilon$ for
$\Gamma < \Gamma_c$, while for $\Gamma \gtrsim \Gamma_c$ it starts
from values close to the gap edge (this peculiar low-temperature
behavior will be explained in subsection~\ref{w}). In the temperature
range $T \lesssim \Delta$, $\omega_c$ is a decreasing function of $T$
for all cases where $\epsilon$ is close to the gap edge (i.e., in deep
doublet and in deep singlet phases), while it is non-monotonic or
increasing for $\epsilon \ll \Delta$ (i.e., in the transition range
with Shiba states deep in the gap), see Fig.~\ref{fig4a}. 

\begin{figure}
\centering
\includegraphics[clip,width=0.48\textwidth]{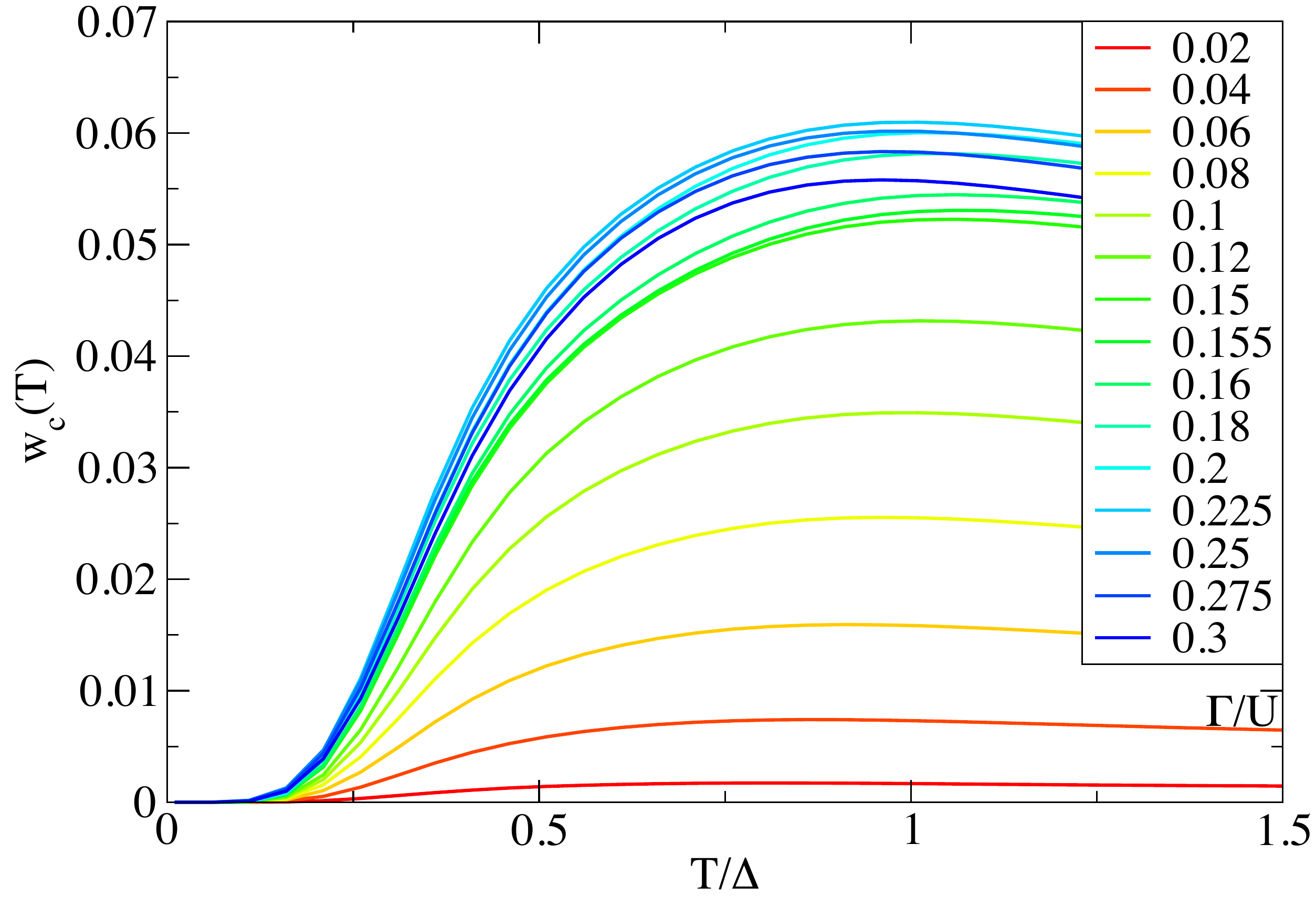}
\includegraphics[clip,width=0.48\textwidth]{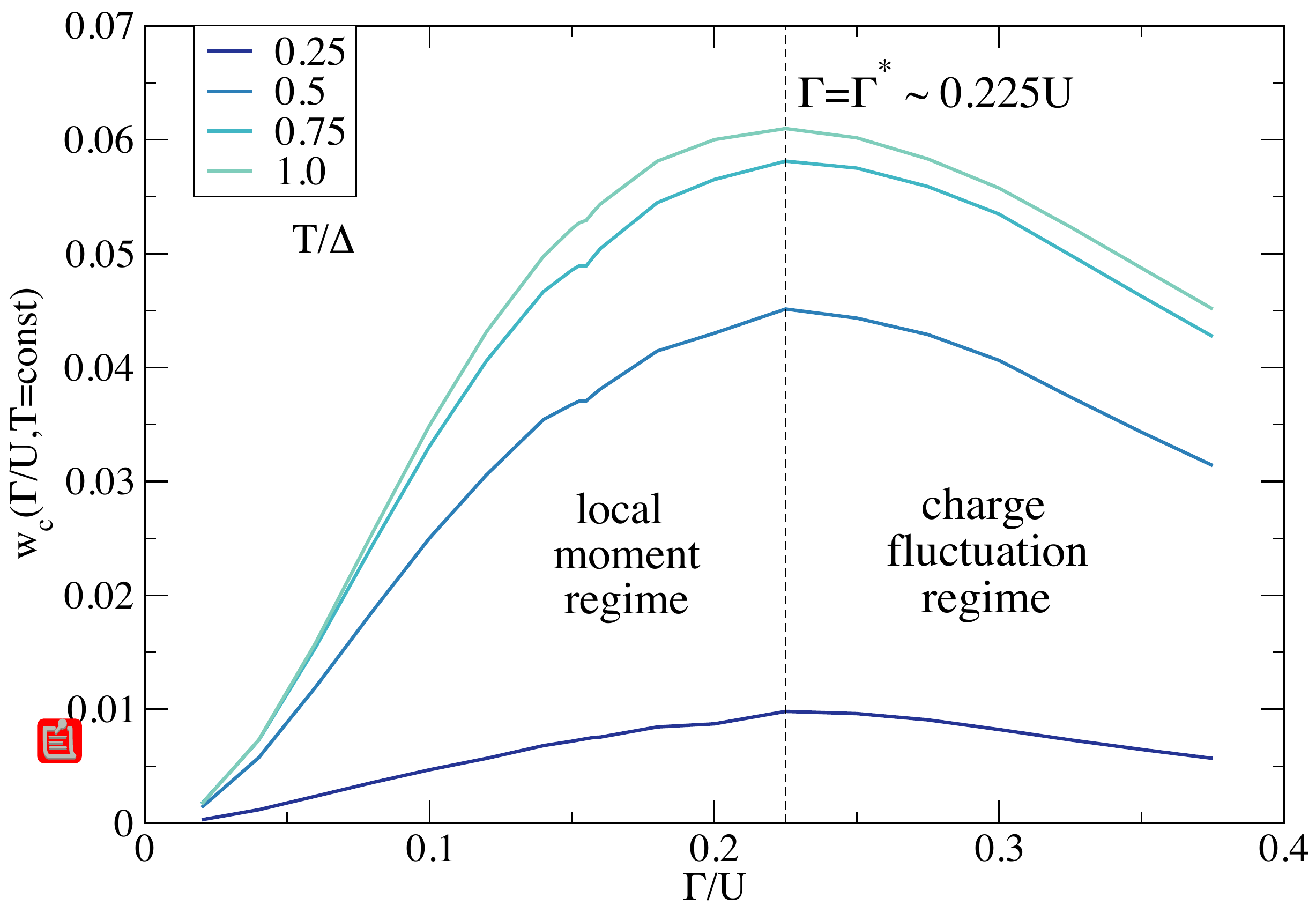}
\caption{(Color online) (a) Temperature dependence of the continuum
part weight, $w_c(T)$, for a range of hybridization strengths
$\Gamma$. (b) $\Gamma$-dependence for a range of fixed temperatures.}
\label{fig4c}
\end{figure}

The continuous-background weight $w_c$ is strictly increasing as a
function of $\Gamma$ at any fixed $T$ up to
\begin{equation}
\Gamma^* \approx 0.225 U,
\end{equation}
see Fig.~\ref{fig4c}. For $\Gamma \lesssim \Gamma^*$, the system is in
the regime of well defined local-moment (the Hartree-Fock solution
spin polarizes for $\Gamma<U/\pi \approx 0.3U$) with properties
controlled by the ratio $\Delta/T_K$, while for $\Gamma \gtrsim
\Gamma^*$ the charge fluctuations are important and the impurity
properties become non-universal. At low $T$, the same exponential law
$w_c = b e^{-\Delta/T}$ is found for all values of $\Gamma \lesssim
\Gamma^*$, both in the singlet and in the doublet regimes, with
$b(\Gamma)$ dependence which can be read off from Fig.~\ref{fig4c}(b).
For $\Gamma \gtrsim \Gamma^*$, however, we find some deviations from
pure exponential dependence. The maximum in $w_c(T)$ is always on the
scale $T \sim \Delta$. 

\begin{figure}
\centering
\includegraphics[clip,width=0.48\textwidth]{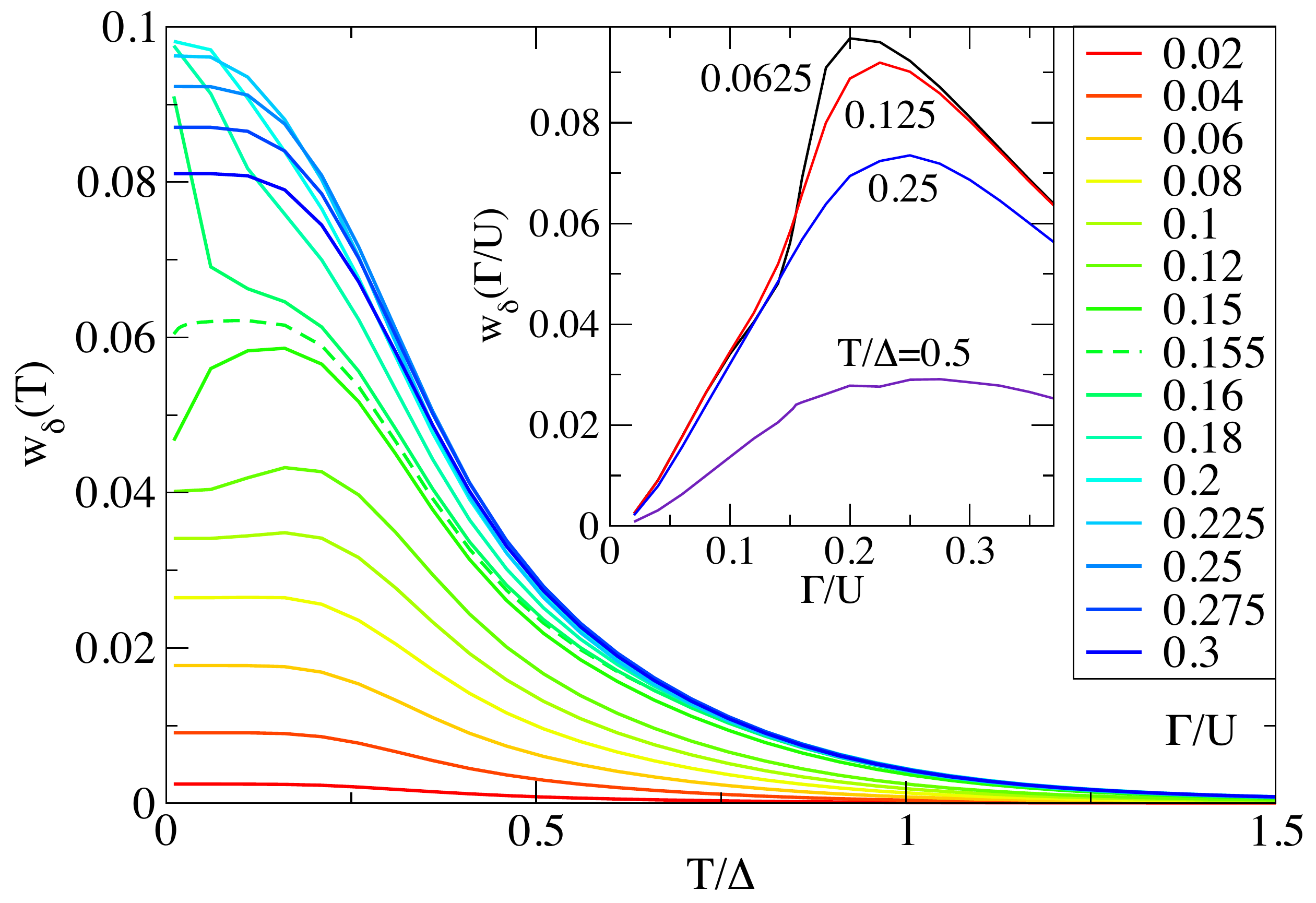}
\caption{(Color online) Temperature dependence of the $\delta$-peak
weight, $w_\delta(T)$, for a range of hybridization strengths
$\Gamma$.}
\label{fig4b}
\end{figure}

The $\delta$-peak weight is monotonically decreasing as a function of
$T$ for small $\Gamma$ and has a local maximum for intermediate
$\Gamma<\Gamma_c$, see Fig.~\ref{fig4b}. The temperature of the
maximum shifts to lower temperatures as $\Gamma$ increases toward
$\Gamma_c$ and for $\Gamma > \Gamma_c$ the weight again becomes a
monotonically decreasing function of $T$. This pronounced difference
in the low-$T$ regime for $\Gamma \approx \Gamma_c$ can serve as a
tool to distinguish between the doublet and singlet regimes at finite
temperatures. Indeed, in the zero-temperature limit and in the absence
of magnetic field (as assumed throughout this work) the sub-gap weight
changes discontinuously by a factor of 2 across the S-D transition,
see the inset to Fig.~\ref{figX}. At finite $T$, this discontinuity is
washed out, see the inset to Fig.~\ref{fig4b}. The up/down-turn of
$w_\delta(T)$ occurs at $T \approx |\epsilon|$, and this scale moves
toward 0 as $\Gamma \to \Gamma_c$, as shown in the main panel of
Fig.~\ref{figX}.

For $\Gamma > \Gamma^*$ the charge fluctuations lead to a
decreasing sub-gap spectral weight. The decreasing trend is also
related to the fact that the $\delta$-peak moves close to the gap edge
in the limit $\Gamma \gg \Gamma^*$. This is a known effect: Shiba
states merge with the continuum in a continuous way by transfering
spectral weight from the $\delta$-peak to the quasiparticle part, so
that the weight of the $\delta$-peak goes to zero as its position
approaches $\omega=\Delta$.

\subsection{High-order Shiba states for large $\Gamma$}
\label{w}

\begin{figure}
\centering
\includegraphics[clip,width=0.48\textwidth]{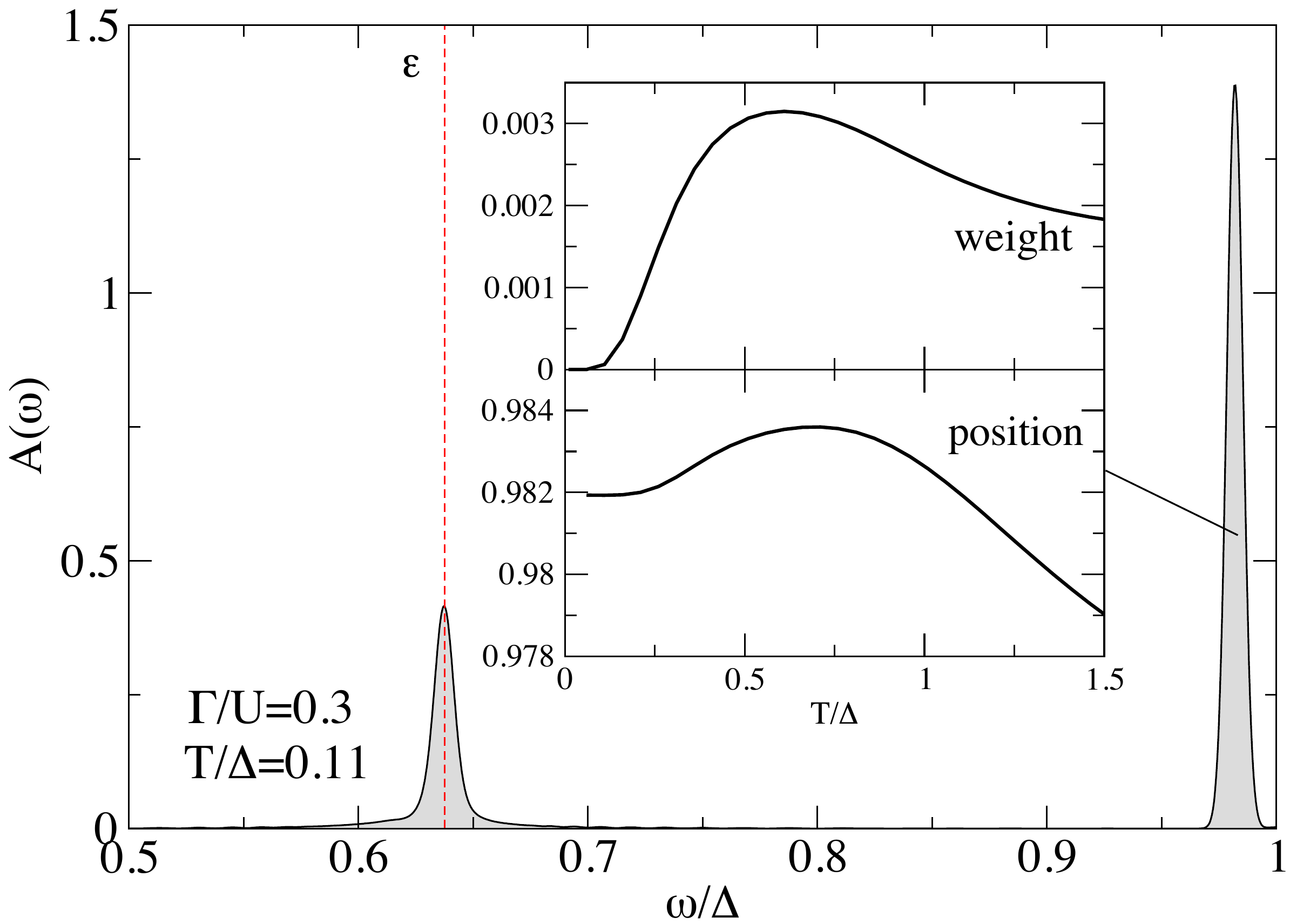}
\caption{(Color online) Sub-gap spectrum for strong hybridization
$\Gamma = 0.3 U$. The inset shows the temperature dependence of the
weight and position of the secondary ``high-order Shiba'' peak which
appears just below the gap edge.}
\label{weird}
\end{figure}

Several anomalies are observed for large values of $\Gamma$. Their common
origin is an additional sub-gap spectral peak just below the gap edge,
see Fig.~\ref{weird}. The weight of this peak shows
activated behavior at low temperatures:
\begin{equation}
w_2(T) = 0.018 e^{-\epsilon/T},
\end{equation}
where $\epsilon=0.637\Delta$ for the chosen value $\Gamma/U=0.3$. This
peak dominates the continuum background for small $T$, because its
activation energy $\epsilon$ is lower than that ($\Delta$) of the
continuous background centered around the Shiba peak. The dominance of
the extra peak in the $T\to0$ limit explains the strikingly peculiar
low-$T$ behavior of $\omega_c(T)$ in Fig.~\ref{fig4a}. Extensive
testing has been performed to assess if this feature could merely be a
numerical artifact of the NRG method. Varying $\Lambda$, Wilson chain
length, the discretization scheme, the algorithm for computing the
spectral function (naive Lehmann-decomposition approach,
complete-Fock-space, full-density-matrix), and the number of states
kept in the truncation, it was found that this feature persists. It is
thus either a generic artifact of the method for finite $T$ and
$\Delta$ that cannot be eliminated by any parameter choice, or a
real spectral feature of the Anderson impurity model with
superconducting baths. Presently, there is no other theoretical method
to reliably confirm the presence of this peak. However, the spectral
weight appears sufficiently large that it could be detected
experimentally, despite its vicinity to the gap edge.

It should be emphasized that there are no discrete sub-gap
multi-particle states with the energy corresponding to this peak.
Instead, its origin is associated with quasiparticle scattering on the
{\sl thermally} excited doublet Shiba state $\ket{D}$ (for large
$\Gamma$, the ground state is namely $\ket{S}$), generating new bound
states of Bogoliubov quasiparticles. In fact, it can be argued that
the physical mechanism is essentially the same as for the conventional
Shiba states: by thermal occupation of the doublet excited states at
finite temperatures the impurity partially ``remagnetizes'', and its
magnetic moment couples to the superconducting bath via an effective
exchange coupling constant proportional to $J_K w_D$, where $J_K
\propto \Gamma/U$ and $w_D = e^{-\epsilon/T}/Z(T)$ is the average
population in the doublet state. This generates a bound state located
just below the gap edge because the effective coupling is weak. This
picture is certainly oversimplified and fails to explain, for example,
the relatively constant position of the peak as a function of
temperature. Nevertheless, it is interesting that such ``high-order
Shiba states'' can be generated at finite temperatures.

\begin{figure}
\centering
\includegraphics[clip,width=0.48\textwidth]{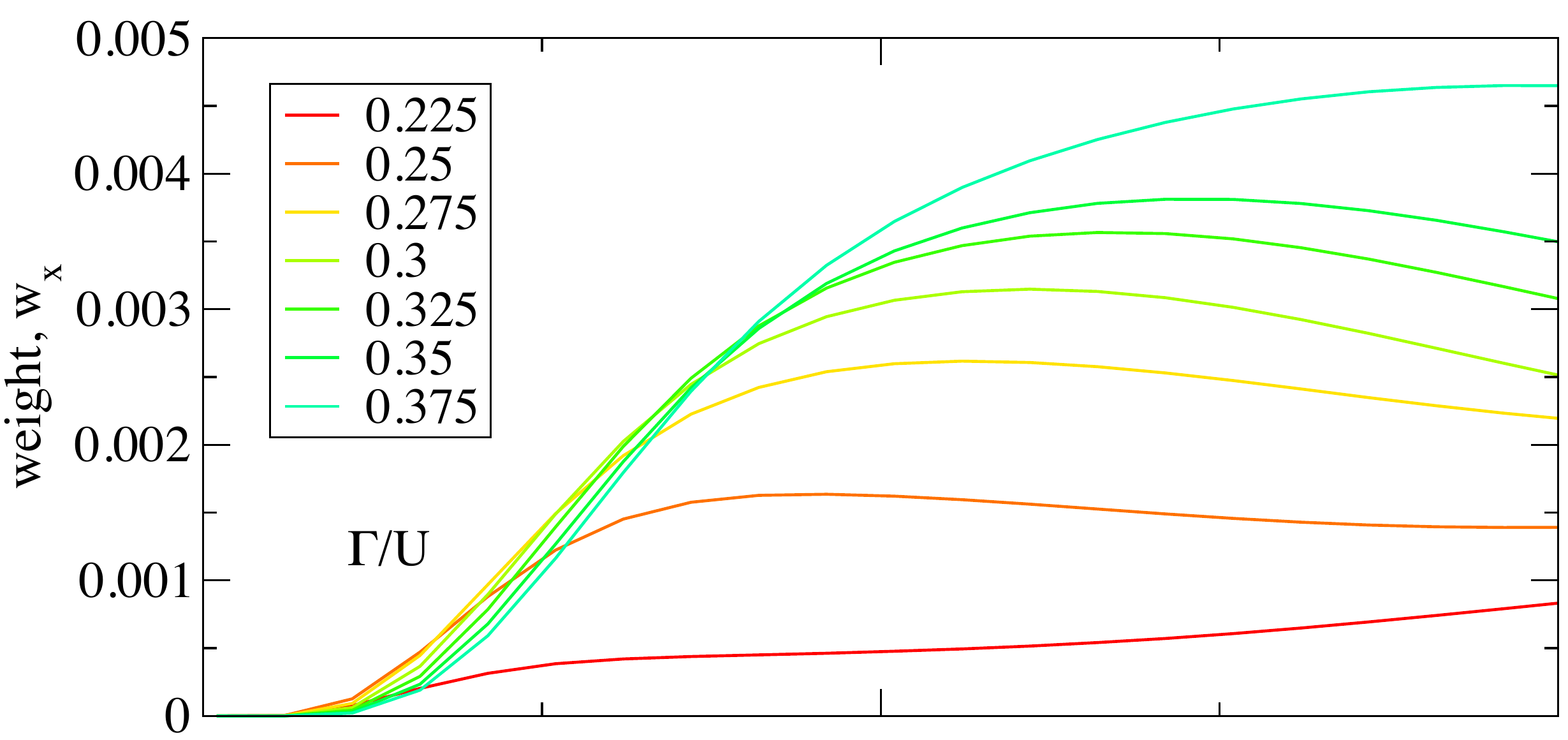}
\includegraphics[clip,width=0.48\textwidth]{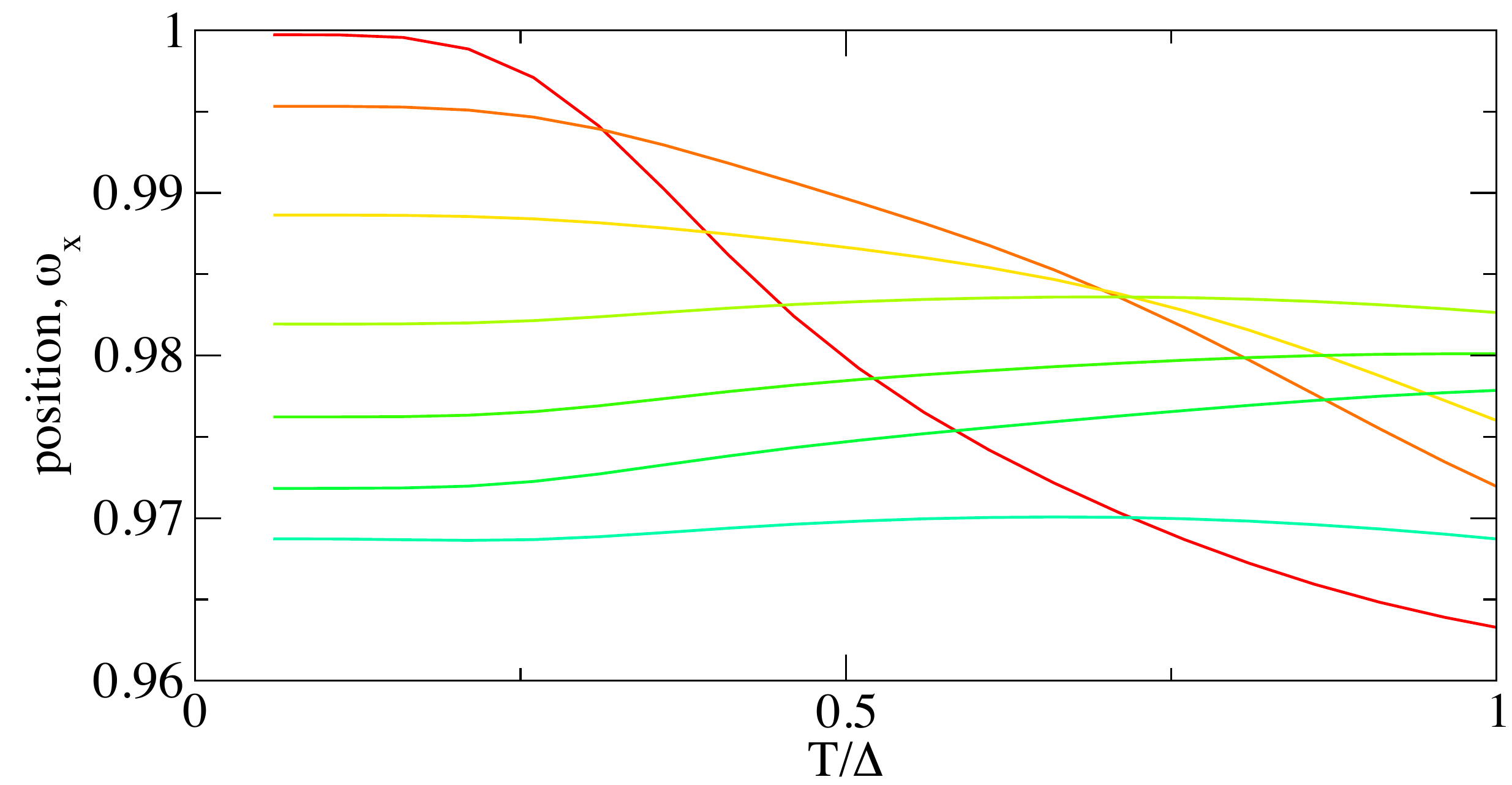}
\caption{(Color online) Weight and position of the ``high-order
Shiba peak'' below the gap edge.}
\label{ww}
\end{figure}

Fig.~\ref{ww} shows the $\Gamma$-dependence of the weight and position
of the additional peak. The threshold for the existence of the peak
is related to $\Gamma^*$, thus the peak is intimately related to
entering the charge-fluctuation regime. Close to the threshold, its
$T=0$ position is at the gap-edge, while for larger $\Gamma$ it starts
at a finite binding energy below the edge.

\section{Results: BCS $\Delta(T)$}
\label{sec4}

We now consider a realistic case where the gap $\Delta$ is temperature
dependent and tends to zero as the critical temperature $T_c$ is
approached. We use a simplified phenomenological expression
\begin{equation}
\Delta_\mathrm{BCS}(T) \approx \delta_{sc} T_c \tanh
\left[
\frac{\pi}{\delta_{sc}} 
\sqrt{a \frac{\delta C}{C_N} \left( \frac{T_c}{T}-1 \right)}
 \,\right]
\end{equation}
with $\delta_{sc}=1.76, a=2/3, \delta C/C_N=1.43$, which is a good
approximation for the true BCS temperature dependence with correct $T\to 0$ and $T\to T_c$ asymptotics.

\begin{figure}
\includegraphics[clip,width=0.48\textwidth]{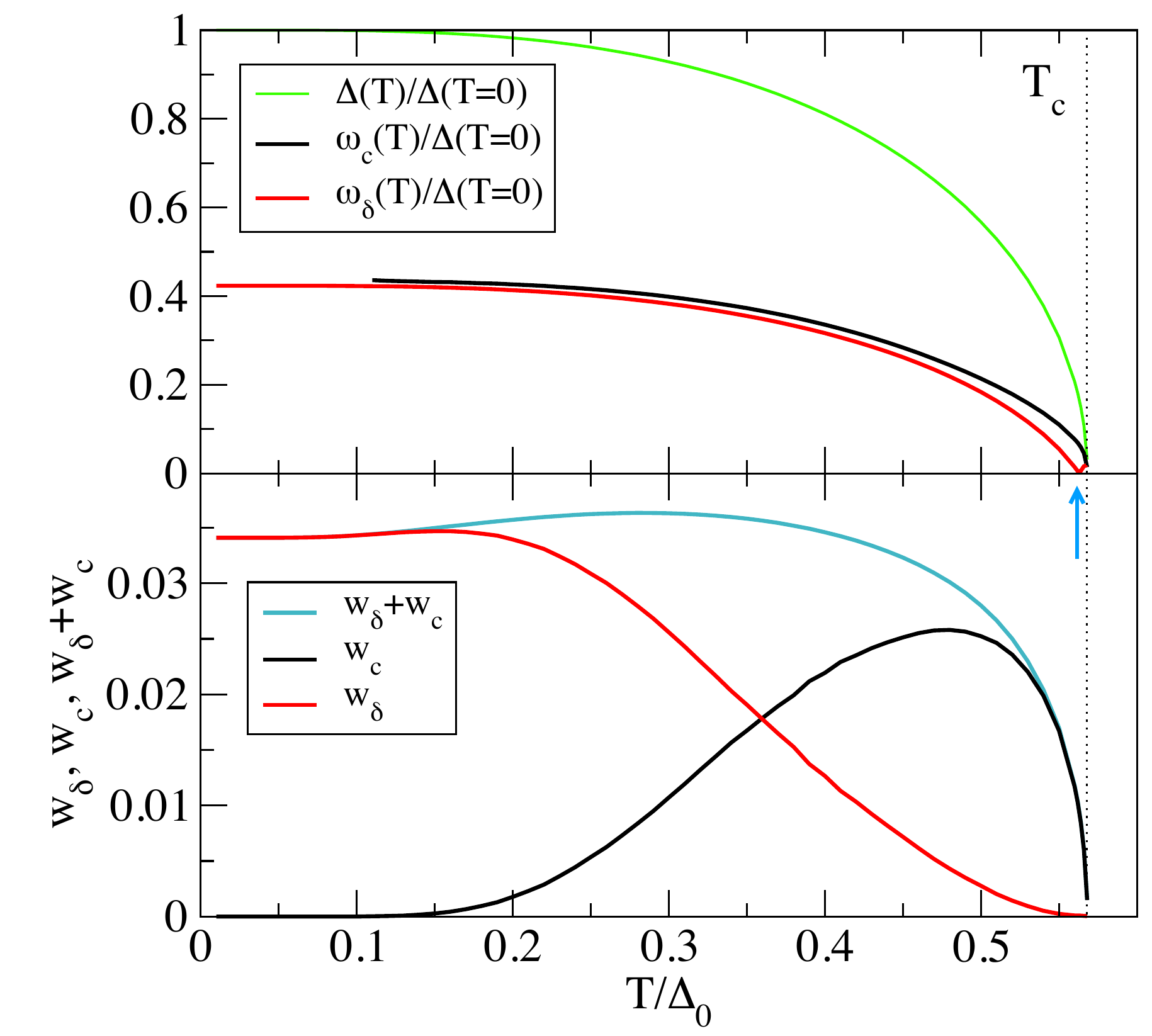}
\caption{(Color online) Temperature dependence of the quantities
characterizing the sub-gap spectral function in the case of
$\Delta=\Delta_\mathrm{BCS}(T)$. The temperature driven
doublet-singlet phase transition is indicated by the arrow. Model
parameters are $\Gamma/U=0.1$ and $U/\Delta=20$.}
\label{k}
\end{figure}

We consider the case where the system is in the doublet regime at
$T=0$. The temperature dependence of key quantities is shown in
Fig.~\ref{k}. The reduction of $\Delta$ with increasing $T$ drives the
system toward the singlet regime. The doublet-singlet transition
occurs, however, just before the critical point (indicated by the
arrow in the figure). Although occurring at a finite temperature, such
first-order boundary transition corresponds to a change of the ground
state of the impurity+bath system by the variation of an ``external''
parameter, thus it may still be considered as a quantum phase
transition of the impurity subsystem (formed by the impurity itself
and the subset of host states which hybridize with the impurity), even
though it is actually driven by thermal fluctuations in the
superconducting host which drive down the gap function $\Delta(T)$.

\section{Discussion}
\label{sec5}

Based on general considerations of an interacting impurity system, and
confirmed by numerical calculations, Shiba states at finite
temperature lose spectral weight to a continuous sub-gap background
centered at the same position. This immediately leads to a question of
principle about the proper definition of the intrinsic lifetime of a
Shiba state. A discrete excited many-particle state isolated from the
continuum could be expected to not decay at all. This is clearly the
case in the absence of quasiparticles. In an open system at finite
temperature, i.e., in contact with a heat and particle reservoar, a
quasiparticle in the superconductor can be generated through a thermal
fluctuation and can interact with the impurity spin, giving rise to a
continuum background. The excited Shiba state can release its
excitation energy to the quasiparticle and decay to the Shiba ground
state, resulting in a finite lifetime. 

The model system studied here is admittedly simplistic. In relastic
systems, in particular when there are tunneling path-ways to a normal
metal (such as a normal-state tip of a scanning tunneling microscope),
the $\delta$-peak will strictly speaking no longer exist. Similarly,
(direct or indirect) coupling to the acoustic phonons of the host will
broaden the $\delta$-peak. If such couplings are small, however, it
may still be expected that the impurity spectral function will be
multimodal with non-trivial temperature dependence.

Let us now consider the example of Mn adatoms on Pb(111) studied in
Ref.~\onlinecite{ruby2015}. Pb has $T_c=\unit[7.2]{K}$ or $\Delta_0
\approx \unit[1.1]{meV}$. The experimental temperatures were
$\unit[1.2]{K}$ and $\unit[4.8]{K}$. Taking into account the reduction
of $\Delta$ in the BCS theory, these correspond to $k_B T/\Delta$ of
$0.12$ and $0.41$, respectively. The lower temperature is thus in the
low-$T$ limit, while at the second one the finite-temperature effects
are expected to be sizable. In experiments, at the lower temperature
the measured linewidth was resolution limited and had to be estimated
indirectly through current saturation plateaus. At the higher
temperature, the width could be extracted from the peak width in the
weak-coupling regime, giving $\Gamma \approx \unit[0.2]{meV}$. Thus
$\Gamma/\Delta_0 \approx 0.2$. Even without discussing how to properly
quantify the intrinsic lifetime in NRG calculations (lower bound is
the HWHM of the continuum peak, $\sim 0.01\Delta$, upper bound is the
standard deviation of the continuum, $\sim 0.1\Delta$), it is possible
to conclude that the order of magnitude is roughly correct. It should
be noted that there are further relaxation mechanisms (such as
fermion-parity-conservig transitions assisted by phonons and photons)
not included in our model, which are likely to be comparable to the
``intrinsic'' broadening due to electron-electron interactions, but
the intrinsic mechanism is certainly not negligible.

This work opens up a number of interesting issues for further study.
One could study how the intrinsic temperature dependence of the
spectral function is reflected in the transport properties. This is
relevant for scanning tunneling spectroscopy studies of single
impurities and adaton chains, such as those expected to host Majorana
end modes. Another question is how the results are modified if the BCS
mean-field Hamiltonian is replaced by a proper interacting model with
electron-electron attraction terms. Finally, we need better
theoretical understanding of the ``high-order Shiba states'' and their
relation to the charge fluctuations.

\begin{acknowledgments}
I acknowledge discussions with Toma\v{z} Rejec and Jernej Mravlje,
and the support of the
Slovenian Research Agency (ARRS) under Program No. P1-0044.
\end{acknowledgments}

\bibliography{paper}

\begin{thebibliography}{29}%
\makeatletter
\providecommand \@ifxundefined [1]{%
 \@ifx{#1\undefined}
}%
\providecommand \@ifnum [1]{%
 \ifnum #1\expandafter \@firstoftwo
 \else \expandafter \@secondoftwo
 \fi
}%
\providecommand \@ifx [1]{%
 \ifx #1\expandafter \@firstoftwo
 \else \expandafter \@secondoftwo
 \fi
}%
\providecommand \natexlab [1]{#1}%
\providecommand \enquote  [1]{``#1''}%
\providecommand \bibnamefont  [1]{#1}%
\providecommand \bibfnamefont [1]{#1}%
\providecommand \citenamefont [1]{#1}%
\providecommand \href@noop [0]{\@secondoftwo}%
\providecommand \href [0]{\begingroup \@sanitize@url \@href}%
\providecommand \@href[1]{\@@startlink{#1}\@@href}%
\providecommand \@@href[1]{\endgroup#1\@@endlink}%
\providecommand \@sanitize@url [0]{\catcode `\\12\catcode `\$12\catcode
  `\&12\catcode `\#12\catcode `\^12\catcode `\_12\catcode `\%12\relax}%
\providecommand \@@startlink[1]{}%
\providecommand \@@endlink[0]{}%
\providecommand \url  [0]{\begingroup\@sanitize@url \@url }%
\providecommand \@url [1]{\endgroup\@href {#1}{\urlprefix }}%
\providecommand \urlprefix  [0]{URL }%
\providecommand \Eprint [0]{\href }%
\providecommand \doibase [0]{http://dx.doi.org/}%
\providecommand \selectlanguage [0]{\@gobble}%
\providecommand \bibinfo  [0]{\@secondoftwo}%
\providecommand \bibfield  [0]{\@secondoftwo}%
\providecommand \translation [1]{[#1]}%
\providecommand \BibitemOpen [0]{}%
\providecommand \bibitemStop [0]{}%
\providecommand \bibitemNoStop [0]{.\EOS\space}%
\providecommand \EOS [0]{\spacefactor3000\relax}%
\providecommand \BibitemShut  [1]{\csname bibitem#1\endcsname}%
\let\auto@bib@innerbib\@empty
%</preamble>
\bibitem [{\citenamefont {Shiba}(1968)}]{shiba1968}%
  \BibitemOpen
  \bibfield  {author} {\bibinfo {author} {\bibfnamefont {H.}~\bibnamefont
  {Shiba}},\ }\bibfield  {title} {\enquote {\bibinfo {title} {Classical spins
  in superconductors},}\ }\href@noop {} {\bibfield  {journal} {\bibinfo
  {journal} {Prog. Theor. Phys.}\ }\textbf {\bibinfo {volume} {40}},\ \bibinfo
  {pages} {435} (\bibinfo {year} {1968})}\BibitemShut {NoStop}%
\bibitem [{\citenamefont {Shiba}(1973)}]{shiba1973}%
  \BibitemOpen
  \bibfield  {author} {\bibinfo {author} {\bibfnamefont {H.}~\bibnamefont
  {Shiba}},\ }\bibfield  {title} {\enquote {\bibinfo {title} {A {Hartree-Fock}
  theory of transition-metal impurities in a superconductor},}\ }\href@noop {}
  {\bibfield  {journal} {\bibinfo  {journal} {Prog. Theor. Phys.}\ }\textbf
  {\bibinfo {volume} {50}},\ \bibinfo {pages} {50} (\bibinfo {year}
  {1973})}\BibitemShut {NoStop}%
\bibitem [{\citenamefont {Satori}\ \emph {et~al.}(1992)\citenamefont {Satori},
  \citenamefont {Shiba}, \citenamefont {Sakai},\ and\ \citenamefont
  {Shimizu}}]{satori1992}%
  \BibitemOpen
  \bibfield  {author} {\bibinfo {author} {\bibfnamefont {Koji}\ \bibnamefont
  {Satori}}, \bibinfo {author} {\bibfnamefont {Hiroyuki}\ \bibnamefont
  {Shiba}}, \bibinfo {author} {\bibfnamefont {Osamu}\ \bibnamefont {Sakai}}, \
  and\ \bibinfo {author} {\bibfnamefont {Yukihiro}\ \bibnamefont {Shimizu}},\
  }\bibfield  {title} {\enquote {\bibinfo {title} {Numerical renormalization
  group study of magnetic impurities in superconductors},}\ }\href@noop {}
  {\bibfield  {journal} {\bibinfo  {journal} {J. Phys. Soc. Japan}\ }\textbf
  {\bibinfo {volume} {61}},\ \bibinfo {pages} {3239} (\bibinfo {year}
  {1992})}\BibitemShut {NoStop}%
\bibitem [{\citenamefont {Balatsky}\ \emph {et~al.}(2006)\citenamefont
  {Balatsky}, \citenamefont {Vekhter},\ and\ \citenamefont
  {Zhu}}]{balatsky2006}%
  \BibitemOpen
  \bibfield  {author} {\bibinfo {author} {\bibfnamefont {A.~V.}\ \bibnamefont
  {Balatsky}}, \bibinfo {author} {\bibfnamefont {I.}~\bibnamefont {Vekhter}}, \
  and\ \bibinfo {author} {\bibfnamefont {Jian-Xin}\ \bibnamefont {Zhu}},\
  }\bibfield  {title} {\enquote {\bibinfo {title} {Impurity-induced states in
  conventional and unconventional superconductors},}\ }\href@noop {} {\bibfield
   {journal} {\bibinfo  {journal} {Rev. Mod. Phys.}\ }\textbf {\bibinfo
  {volume} {78}},\ \bibinfo {pages} {373} (\bibinfo {year} {2006})}\BibitemShut
  {NoStop}%
\bibitem [{\citenamefont {Yazdani}\ \emph {et~al.}(1997)\citenamefont
  {Yazdani}, \citenamefont {Jones}, \citenamefont {Lutz}, \citenamefont
  {Crommie},\ and\ \citenamefont {Eigler}}]{yazdani1997}%
  \BibitemOpen
  \bibfield  {author} {\bibinfo {author} {\bibfnamefont {A.}~\bibnamefont
  {Yazdani}}, \bibinfo {author} {\bibfnamefont {B.~A.}\ \bibnamefont {Jones}},
  \bibinfo {author} {\bibfnamefont {C.~P.}\ \bibnamefont {Lutz}}, \bibinfo
  {author} {\bibfnamefont {M.~F.}\ \bibnamefont {Crommie}}, \ and\ \bibinfo
  {author} {\bibfnamefont {D.~M.}\ \bibnamefont {Eigler}},\ }\bibfield  {title}
  {\enquote {\bibinfo {title} {Probing the local effects of magnetic impurities
  on superconductivity},}\ }\href@noop {} {\bibfield  {journal} {\bibinfo
  {journal} {Science}\ }\textbf {\bibinfo {volume} {275}},\ \bibinfo {pages}
  {1767} (\bibinfo {year} {1997})}\BibitemShut {NoStop}%
\bibitem [{\citenamefont {Deacon}\ \emph {et~al.}(2010)\citenamefont {Deacon},
  \citenamefont {Tanaka}, \citenamefont {Oiwa}, \citenamefont {Sakano},
  \citenamefont {Yoshida}, \citenamefont {Shibata}, \citenamefont {Hirakawa},\
  and\ \citenamefont {Tarucha}}]{Deacon:2010jn}%
  \BibitemOpen
  \bibfield  {author} {\bibinfo {author} {\bibfnamefont {R~S}\ \bibnamefont
  {Deacon}}, \bibinfo {author} {\bibfnamefont {Y}~\bibnamefont {Tanaka}},
  \bibinfo {author} {\bibfnamefont {A}~\bibnamefont {Oiwa}}, \bibinfo {author}
  {\bibfnamefont {R}~\bibnamefont {Sakano}}, \bibinfo {author} {\bibfnamefont
  {K}~\bibnamefont {Yoshida}}, \bibinfo {author} {\bibfnamefont
  {K}~\bibnamefont {Shibata}}, \bibinfo {author} {\bibfnamefont
  {K}~\bibnamefont {Hirakawa}}, \ and\ \bibinfo {author} {\bibfnamefont
  {S}~\bibnamefont {Tarucha}},\ }\bibfield  {title} {\enquote {\bibinfo {title}
  {{Interplay of Kondo and superconducting correlations in the nonequilibrium
  Andreev transport through a quantum dot}},}\ }\href@noop {} {\bibfield
  {journal} {\bibinfo  {journal} {Physical Review Letters}\ }\textbf {\bibinfo
  {volume} {104}},\ \bibinfo {pages} {076805} (\bibinfo {year}
  {2010})}\BibitemShut {NoStop}%
\bibitem [{\citenamefont {Pillet}\ \emph {et~al.}(2010)\citenamefont {Pillet},
  \citenamefont {Quay}, \citenamefont {Morin}, \citenamefont {Bena},
  \citenamefont {Yeyati},\ and\ \citenamefont {Joyez}}]{pillet2010}%
  \BibitemOpen
  \bibfield  {author} {\bibinfo {author} {\bibfnamefont {J.-D.}\ \bibnamefont
  {Pillet}}, \bibinfo {author} {\bibfnamefont {C.~H.~L.}\ \bibnamefont {Quay}},
  \bibinfo {author} {\bibfnamefont {P.}~\bibnamefont {Morin}}, \bibinfo
  {author} {\bibfnamefont {C.}~\bibnamefont {Bena}}, \bibinfo {author}
  {\bibfnamefont {A.~Levy}\ \bibnamefont {Yeyati}}, \ and\ \bibinfo {author}
  {\bibfnamefont {P.}~\bibnamefont {Joyez}},\ }\bibfield  {title} {\enquote
  {\bibinfo {title} {Andreev bound states in supercurrent-carrying carbon
  nanotubes revealed},}\ }\href@noop {} {\bibfield  {journal} {\bibinfo
  {journal} {Nat. Physics}\ }\textbf {\bibinfo {volume} {6}},\ \bibinfo {pages}
  {965} (\bibinfo {year} {2010})}\BibitemShut {NoStop}%
\bibitem [{\citenamefont {Franke}\ \emph {et~al.}(2011)\citenamefont {Franke},
  \citenamefont {Schulze},\ and\ \citenamefont {Pascual}}]{franke2011}%
  \BibitemOpen
  \bibfield  {author} {\bibinfo {author} {\bibfnamefont {K.~J.}\ \bibnamefont
  {Franke}}, \bibinfo {author} {\bibfnamefont {G.}~\bibnamefont {Schulze}}, \
  and\ \bibinfo {author} {\bibfnamefont {J.~I.}\ \bibnamefont {Pascual}},\
  }\bibfield  {title} {\enquote {\bibinfo {title} {Competition of
  superconductivity phenomena and {Kondo} screening at the nanoscale},}\
  }\href@noop {} {\bibfield  {journal} {\bibinfo  {journal} {Science}\ }\textbf
  {\bibinfo {volume} {332}},\ \bibinfo {pages} {940} (\bibinfo {year}
  {2011})}\BibitemShut {NoStop}%
\bibitem [{\citenamefont {Mart{\'\i}n-Rodero}\ and\ \citenamefont
  {Levy~Yeyati}(2011)}]{rodero2011}%
  \BibitemOpen
  \bibfield  {author} {\bibinfo {author} {\bibfnamefont {A.}~\bibnamefont
  {Mart{\'\i}n-Rodero}}\ and\ \bibinfo {author} {\bibfnamefont
  {A.}~\bibnamefont {Levy~Yeyati}},\ }\bibfield  {title} {\enquote {\bibinfo
  {title} {Josephson and {Andreev} transport through quantum dots},}\
  }\href@noop {} {\bibfield  {journal} {\bibinfo  {journal} {Advances in
  Physics}\ }\textbf {\bibinfo {volume} {60}},\ \bibinfo {pages} {899--958}
  (\bibinfo {year} {2011})}\BibitemShut {NoStop}%
\bibitem [{\citenamefont {Kumar}\ \emph {et~al.}(2014)\citenamefont {Kumar},
  \citenamefont {Gaim}, \citenamefont {Steininger}, \citenamefont {Yeyati},
  \citenamefont {Mart{\'\i}n-Rodero}, \citenamefont {H{\"u}ttel},\ and\
  \citenamefont {Strunk}}]{Kumar:2014cq}%
  \BibitemOpen
  \bibfield  {author} {\bibinfo {author} {\bibfnamefont {A}~\bibnamefont
  {Kumar}}, \bibinfo {author} {\bibfnamefont {M}~\bibnamefont {Gaim}}, \bibinfo
  {author} {\bibfnamefont {D}~\bibnamefont {Steininger}}, \bibinfo {author}
  {\bibfnamefont {A~Levy}\ \bibnamefont {Yeyati}}, \bibinfo {author}
  {\bibfnamefont {A}~\bibnamefont {Mart{\'\i}n-Rodero}}, \bibinfo {author}
  {\bibfnamefont {A~K}\ \bibnamefont {H{\"u}ttel}}, \ and\ \bibinfo {author}
  {\bibfnamefont {C}~\bibnamefont {Strunk}},\ }\bibfield  {title} {\enquote
  {\bibinfo {title} {{Temperature dependence of Andreev spectra in a
  superconducting carbon nanotube quantum dot}},}\ }\href@noop {} {\bibfield
  {journal} {\bibinfo  {journal} {Physical Review B}\ }\textbf {\bibinfo
  {volume} {89}},\ \bibinfo {pages} {075428} (\bibinfo {year}
  {2014})}\BibitemShut {NoStop}%
\bibitem [{\citenamefont {Ruby}\ \emph {et~al.}(2015)\citenamefont {Ruby},
  \citenamefont {Pientka}, \citenamefont {Peng}, \citenamefont {von Oppen},
  \citenamefont {Heinrich},\ and\ \citenamefont {Franke}}]{ruby2015}%
  \BibitemOpen
  \bibfield  {author} {\bibinfo {author} {\bibfnamefont {Michael}\ \bibnamefont
  {Ruby}}, \bibinfo {author} {\bibfnamefont {Falko}\ \bibnamefont {Pientka}},
  \bibinfo {author} {\bibfnamefont {Yang}\ \bibnamefont {Peng}}, \bibinfo
  {author} {\bibfnamefont {Felix}\ \bibnamefont {von Oppen}}, \bibinfo {author}
  {\bibfnamefont {Benjamin~W}\ \bibnamefont {Heinrich}}, \ and\ \bibinfo
  {author} {\bibfnamefont {Katharina~J}\ \bibnamefont {Franke}},\ }\bibfield
  {title} {\enquote {\bibinfo {title} {{Tunneling Processes into Localized
  Subgap States in Superconductors}},}\ }\href@noop {} {\bibfield  {journal}
  {\bibinfo  {journal} {Physical Review Letters}\ }\textbf {\bibinfo {volume}
  {115}},\ \bibinfo {pages} {087001--5} (\bibinfo {year} {2015})}\BibitemShut
  {NoStop}%
\bibitem [{\citenamefont {Krishna-murthy}\ \emph {et~al.}(1980)\citenamefont
  {Krishna-murthy}, \citenamefont {Wilkins},\ and\ \citenamefont
  {Wilson}}]{krishna1980a}%
  \BibitemOpen
  \bibfield  {author} {\bibinfo {author} {\bibfnamefont {H.~R.}\ \bibnamefont
  {Krishna-murthy}}, \bibinfo {author} {\bibfnamefont {J.~W.}\ \bibnamefont
  {Wilkins}}, \ and\ \bibinfo {author} {\bibfnamefont {K.~G.}\ \bibnamefont
  {Wilson}},\ }\bibfield  {title} {\enquote {\bibinfo {title}
  {Renormalization-group approach to the {Anderson} model of dilute magnetic
  alloys. {I.} {S}tatic properties for the symmetric case},}\ }\href@noop {}
  {\bibfield  {journal} {\bibinfo  {journal} {Phys. Rev. B}\ }\textbf {\bibinfo
  {volume} {21}},\ \bibinfo {pages} {1003} (\bibinfo {year}
  {1980})}\BibitemShut {NoStop}%
\bibitem [{\citenamefont {Wilson}(1975)}]{wilson1975}%
  \BibitemOpen
  \bibfield  {author} {\bibinfo {author} {\bibfnamefont {K.~G.}\ \bibnamefont
  {Wilson}},\ }\bibfield  {title} {\enquote {\bibinfo {title} {The
  renormalization group: {Critical} phenomena and the {Kondo} problem},}\
  }\href@noop {} {\bibfield  {journal} {\bibinfo  {journal} {Rev. Mod. Phys.}\
  }\textbf {\bibinfo {volume} {47}},\ \bibinfo {pages} {773} (\bibinfo {year}
  {1975})}\BibitemShut {NoStop}%
\bibitem [{\citenamefont {Sakai}\ \emph {et~al.}(1993)\citenamefont {Sakai},
  \citenamefont {Shimizu}, \citenamefont {Shiba},\ and\ \citenamefont
  {Satori}}]{sakai1993}%
  \BibitemOpen
  \bibfield  {author} {\bibinfo {author} {\bibfnamefont {Osamu}\ \bibnamefont
  {Sakai}}, \bibinfo {author} {\bibfnamefont {Yukihiro}\ \bibnamefont
  {Shimizu}}, \bibinfo {author} {\bibfnamefont {Hiroyuki}\ \bibnamefont
  {Shiba}}, \ and\ \bibinfo {author} {\bibfnamefont {Koji}\ \bibnamefont
  {Satori}},\ }\bibfield  {title} {\enquote {\bibinfo {title} {Numerical
  renormalization group study of magnetic impurities in supercoductors. {II.}
  {D}ynamical excitations spectra and spatial variation of the order
  parameter},}\ }\href@noop {} {\bibfield  {journal} {\bibinfo  {journal} {J.
  Phys. Soc. Japan}\ }\textbf {\bibinfo {volume} {62}},\ \bibinfo {pages}
  {3181} (\bibinfo {year} {1993})}\BibitemShut {NoStop}%
\bibitem [{\citenamefont {Yoshioka}\ and\ \citenamefont
  {Ohashi}(2000)}]{yoshioka2000}%
  \BibitemOpen
  \bibfield  {author} {\bibinfo {author} {\bibfnamefont {Tomoki}\ \bibnamefont
  {Yoshioka}}\ and\ \bibinfo {author} {\bibfnamefont {Yoji}\ \bibnamefont
  {Ohashi}},\ }\bibfield  {title} {\enquote {\bibinfo {title} {Numerical
  renormalization group studies on single impurity anderson model in
  superconductivity: a unified treatment of magnetic, nonmagnetic impurities,
  and resonance scattering},}\ }\href@noop {} {\bibfield  {journal} {\bibinfo
  {journal} {J. Phys. Soc. Japan}\ }\textbf {\bibinfo {volume} {69}},\ \bibinfo
  {pages} {1812} (\bibinfo {year} {2000})}\BibitemShut {NoStop}%
\bibitem [{\citenamefont {Oguri}\ \emph {et~al.}(2004)\citenamefont {Oguri},
  \citenamefont {Tanaka},\ and\ \citenamefont {Hewson}}]{oguri2004josephson}%
  \BibitemOpen
  \bibfield  {author} {\bibinfo {author} {\bibfnamefont {Akira}\ \bibnamefont
  {Oguri}}, \bibinfo {author} {\bibfnamefont {Yoshihide}\ \bibnamefont
  {Tanaka}}, \ and\ \bibinfo {author} {\bibfnamefont {A.~C.}\ \bibnamefont
  {Hewson}},\ }\bibfield  {title} {\enquote {\bibinfo {title} {Quantum phase
  transition in a minimal model for the {Kondo} effect in a {Josephson}
  junction},}\ }\href@noop {} {\bibfield  {journal} {\bibinfo  {journal} {J.
  Phys. Soc. Japan}\ }\textbf {\bibinfo {volume} {73}},\ \bibinfo {pages}
  {2494} (\bibinfo {year} {2004})}\BibitemShut {NoStop}%
\bibitem [{\citenamefont {Bauer}\ \emph {et~al.}(2007)\citenamefont {Bauer},
  \citenamefont {Oguri},\ and\ \citenamefont {Hewson}}]{bauer2007}%
  \BibitemOpen
  \bibfield  {author} {\bibinfo {author} {\bibfnamefont {J.}~\bibnamefont
  {Bauer}}, \bibinfo {author} {\bibfnamefont {A.}~\bibnamefont {Oguri}}, \ and\
  \bibinfo {author} {\bibfnamefont {A.~C.}\ \bibnamefont {Hewson}},\ }\bibfield
   {title} {\enquote {\bibinfo {title} {Spectral properties of locally
  correlated electrons in a {Bardeen-Cooper-Schrieffer} superconductor},}\
  }\href@noop {} {\bibfield  {journal} {\bibinfo  {journal} {J. Phys.: Condens.
  Matter}\ }\textbf {\bibinfo {volume} {19}},\ \bibinfo {pages} {486211}
  (\bibinfo {year} {2007})}\BibitemShut {NoStop}%
\bibitem [{\citenamefont {Karrasch}\ \emph {et~al.}(2008)\citenamefont
  {Karrasch}, \citenamefont {Oguri},\ and\ \citenamefont
  {Meden}}]{karrasch2008}%
  \BibitemOpen
  \bibfield  {author} {\bibinfo {author} {\bibfnamefont {C.}~\bibnamefont
  {Karrasch}}, \bibinfo {author} {\bibfnamefont {A.}~\bibnamefont {Oguri}}, \
  and\ \bibinfo {author} {\bibfnamefont {V.}~\bibnamefont {Meden}},\ }\bibfield
   {title} {\enquote {\bibinfo {title} {Josephson current through a single
  {Anderson} impurity coupled to {BCS} leads},}\ }\href@noop {} {\bibfield
  {journal} {\bibinfo  {journal} {Phys. Rev. B}\ }\textbf {\bibinfo {volume}
  {77}},\ \bibinfo {pages} {024517} (\bibinfo {year} {2008})}\BibitemShut
  {NoStop}%
\bibitem [{\citenamefont {Bulla}\ \emph {et~al.}(2008)\citenamefont {Bulla},
  \citenamefont {Costi},\ and\ \citenamefont {Pruschke}}]{bulla2008}%
  \BibitemOpen
  \bibfield  {author} {\bibinfo {author} {\bibfnamefont {Ralf}\ \bibnamefont
  {Bulla}}, \bibinfo {author} {\bibfnamefont {Theo}\ \bibnamefont {Costi}}, \
  and\ \bibinfo {author} {\bibfnamefont {Thomas}\ \bibnamefont {Pruschke}},\
  }\bibfield  {title} {\enquote {\bibinfo {title} {The numerical
  renormalization group method for quantum impurity systems},}\ }\href@noop {}
  {\bibfield  {journal} {\bibinfo  {journal} {Rev. Mod. Phys.}\ }\textbf
  {\bibinfo {volume} {80}},\ \bibinfo {pages} {395} (\bibinfo {year}
  {2008})}\BibitemShut {NoStop}%
\bibitem [{\citenamefont {\v{Z}itko}\ \emph {et~al.}(2015)\citenamefont
  {\v{Z}itko}, \citenamefont {Lim}, \citenamefont {Lopez},\ and\ \citenamefont
  {Aguado}}]{zitko2015shiba}%
  \BibitemOpen
  \bibfield  {author} {\bibinfo {author} {\bibfnamefont {R.}~\bibnamefont
  {\v{Z}itko}}, \bibinfo {author} {\bibfnamefont {Jong~Soo}\ \bibnamefont
  {Lim}}, \bibinfo {author} {\bibfnamefont {Rosa}\ \bibnamefont {Lopez}}, \
  and\ \bibinfo {author} {\bibfnamefont {Ramon}\ \bibnamefont {Aguado}},\
  }\bibfield  {title} {\enquote {\bibinfo {title} {Shiba states and zero-bias
  anomalies in the hybrid normal-superconductor {Anderson} model},}\
  }\href@noop {} {\bibfield  {journal} {\bibinfo  {journal} {Phys. Rev. B}\
  }\textbf {\bibinfo {volume} {91}},\ \bibinfo {pages} {045441} (\bibinfo
  {year} {2015})}\BibitemShut {NoStop}%
\bibitem [{\citenamefont {Luitz}\ and\ \citenamefont
  {Assaad}(2010)}]{Luitz:2010bn}%
  \BibitemOpen
  \bibfield  {author} {\bibinfo {author} {\bibfnamefont {David~J}\ \bibnamefont
  {Luitz}}\ and\ \bibinfo {author} {\bibfnamefont {Fakher~F}\ \bibnamefont
  {Assaad}},\ }\bibfield  {title} {\enquote {\bibinfo {title} {{Weak-coupling
  continuous-time quantum Monte Carlo study of the single impurity and periodic
  Anderson models with s-wave superconducting baths}},}\ }\href@noop {}
  {\bibfield  {journal} {\bibinfo  {journal} {Physical Review B}\ }\textbf
  {\bibinfo {volume} {81}},\ \bibinfo {pages} {024509} (\bibinfo {year}
  {2010})}\BibitemShut {NoStop}%
\bibitem [{\citenamefont {Gull}\ \emph {et~al.}(2011)\citenamefont {Gull},
  \citenamefont {Millis}, \citenamefont {Lichtenstein}, \citenamefont
  {Rubtsov}, \citenamefont {Troyer},\ and\ \citenamefont {Werner}}]{gull2011}%
  \BibitemOpen
  \bibfield  {author} {\bibinfo {author} {\bibfnamefont {E.}~\bibnamefont
  {Gull}}, \bibinfo {author} {\bibfnamefont {A.~J.}\ \bibnamefont {Millis}},
  \bibinfo {author} {\bibfnamefont {A.~I.}\ \bibnamefont {Lichtenstein}},
  \bibinfo {author} {\bibfnamefont {A.~N.}\ \bibnamefont {Rubtsov}}, \bibinfo
  {author} {\bibfnamefont {M.}~\bibnamefont {Troyer}}, \ and\ \bibinfo {author}
  {\bibfnamefont {P.}~\bibnamefont {Werner}},\ }\bibfield  {title} {\enquote
  {\bibinfo {title} {Continuous-time monte carlo methods for quantum impurity
  models},}\ }\href@noop {} {\bibfield  {journal} {\bibinfo  {journal} {Rev.
  Mod. Phys.}\ }\textbf {\bibinfo {volume} {83}},\ \bibinfo {pages} {349}
  (\bibinfo {year} {2011})}\BibitemShut {NoStop}%
\bibitem [{\citenamefont {Oliveira}\ and\ \citenamefont
  {Oliveira}(1994)}]{oliveira1994}%
  \BibitemOpen
  \bibfield  {author} {\bibinfo {author} {\bibfnamefont {W.~C.}\ \bibnamefont
  {Oliveira}}\ and\ \bibinfo {author} {\bibfnamefont {L.~N.}\ \bibnamefont
  {Oliveira}},\ }\bibfield  {title} {\enquote {\bibinfo {title} {Generalized
  numerical renormalization-group method to calculate the thermodynamical
  properties of impurities in metals},}\ }\href@noop {} {\bibfield  {journal}
  {\bibinfo  {journal} {Phys. Rev. B}\ }\textbf {\bibinfo {volume} {49}},\
  \bibinfo {pages} {11986} (\bibinfo {year} {1994})}\BibitemShut {NoStop}%
\bibitem [{\citenamefont {\v{Z}itko}\ and\ \citenamefont
  {Pruschke}(2009)}]{resolution}%
  \BibitemOpen
  \bibfield  {author} {\bibinfo {author} {\bibfnamefont {Rok}\ \bibnamefont
  {\v{Z}itko}}\ and\ \bibinfo {author} {\bibfnamefont {Thomas}\ \bibnamefont
  {Pruschke}},\ }\bibfield  {title} {\enquote {\bibinfo {title} {Energy
  resolution and discretization artefacts in the numerical renormalization
  group},}\ }\href@noop {} {\bibfield  {journal} {\bibinfo  {journal} {Phys.
  Rev. B}\ }\textbf {\bibinfo {volume} {79}},\ \bibinfo {pages} {085106}
  (\bibinfo {year} {2009})}\BibitemShut {NoStop}%
\bibitem [{\citenamefont {Anders}\ and\ \citenamefont
  {Schiller}(2005)}]{anders2005}%
  \BibitemOpen
  \bibfield  {author} {\bibinfo {author} {\bibfnamefont {F.~B.}\ \bibnamefont
  {Anders}}\ and\ \bibinfo {author} {\bibfnamefont {A.}~\bibnamefont
  {Schiller}},\ }\bibfield  {title} {\enquote {\bibinfo {title} {Real-time
  dynamics in quantum impurity systems: A time-dependent numerical
  renormalization group approach},}\ }\href@noop {} {\bibfield  {journal}
  {\bibinfo  {journal} {Phys. Rev. Lett.}\ }\textbf {\bibinfo {volume} {95}},\
  \bibinfo {pages} {196801} (\bibinfo {year} {2005})}\BibitemShut {NoStop}%
\bibitem [{\citenamefont {Peters}\ \emph {et~al.}(2006)\citenamefont {Peters},
  \citenamefont {Pruschke},\ and\ \citenamefont {Anders}}]{peters2006}%
  \BibitemOpen
  \bibfield  {author} {\bibinfo {author} {\bibfnamefont {Robert}\ \bibnamefont
  {Peters}}, \bibinfo {author} {\bibfnamefont {Thomas}\ \bibnamefont
  {Pruschke}}, \ and\ \bibinfo {author} {\bibfnamefont {Frithjof~B.}\
  \bibnamefont {Anders}},\ }\bibfield  {title} {\enquote {\bibinfo {title} {A
  numerical renormalization group approach to {Green's} functions for quantum
  impurity models},}\ }\href@noop {} {\bibfield  {journal} {\bibinfo  {journal}
  {Phys. Rev. B}\ }\textbf {\bibinfo {volume} {74}},\ \bibinfo {pages} {245114}
  (\bibinfo {year} {2006})}\BibitemShut {NoStop}%
\bibitem [{\citenamefont {Weichselbaum}\ and\ \citenamefont {von
  Delft}(2007)}]{weichselbaum2007}%
  \BibitemOpen
  \bibfield  {author} {\bibinfo {author} {\bibfnamefont {Andreas}\ \bibnamefont
  {Weichselbaum}}\ and\ \bibinfo {author} {\bibfnamefont {Jan}\ \bibnamefont
  {von Delft}},\ }\bibfield  {title} {\enquote {\bibinfo {title} {Sum-rule
  conserving spectral functions from the numerical renormalization group},}\
  }\href@noop {} {\bibfield  {journal} {\bibinfo  {journal} {Phys. Rev. Lett.}\
  }\textbf {\bibinfo {volume} {99}},\ \bibinfo {pages} {076402} (\bibinfo
  {year} {2007})}\BibitemShut {NoStop}%
\bibitem [{Note1()}]{Note1}%
  \BibitemOpen
  \bibinfo {note} {A possible improvement consists in formulating the NRG
  truncation rule so that a comparable total number of multiplets is kept in
  even- and odd-fermion-parity sectors, but this has not yet been tried
  out.}\BibitemShut {Stop}%
\bibitem [{\citenamefont {Hecht}\ \emph {et~al.}(2008)\citenamefont {Hecht},
  \citenamefont {Weichselbaum}, \citenamefont {von Delft},\ and\ \citenamefont
  {Bulla}}]{hecht2008}%
  \BibitemOpen
  \bibfield  {author} {\bibinfo {author} {\bibfnamefont {T.}~\bibnamefont
  {Hecht}}, \bibinfo {author} {\bibfnamefont {A.}~\bibnamefont {Weichselbaum}},
  \bibinfo {author} {\bibfnamefont {J.}~\bibnamefont {von Delft}}, \ and\
  \bibinfo {author} {\bibfnamefont {R.}~\bibnamefont {Bulla}},\ }\bibfield
  {title} {\enquote {\bibinfo {title} {Numerical renormalization group
  calculation of near-gap peaks in spectral functions of the {Anderson} model
  with superconducting leads},}\ }\href@noop {} {\bibfield  {journal} {\bibinfo
   {journal} {J. Phys. Condens. Mat.}\ }\textbf {\bibinfo {volume} {20}},\
  \bibinfo {pages} {275213} (\bibinfo {year} {2008})}\BibitemShut {NoStop}%
\end{thebibliography}%

\end{document}